\documentclass[epj]{svjour}
\usepackage{graphicx,epstopdf,multirow,amsmath,rotating,subfigure,isotope,gensymb,amssymb}
\usepackage{graphicx}% Include figure files
\usepackage{dcolumn}% Align table columns on decimal point
\usepackage{bm}% bold math
\usepackage{tabularx}
\usepackage[export]{adjustbox}
\usepackage{float}
\usepackage{braket}
\usepackage{lscape}
\usepackage{lipsum}
\usepackage{longtable}

\usepackage{latexsym}
\usepackage{graphics}
\usepackage{xcolor}   %May be necessary if you want to color links

\usepackage{hyperref}
\hypersetup{
    colorlinks=true, %set true if you want colored links
    linktoc=all,     %set to all if you want both sections and subsections linked
    urlcolor  = blue,
    citecolor = blue,
    linkcolor=blue,  %choose some color if you want links to stand out
}
\makeatletter
\newcommand{\colorcaption}[2][]{%
  \begingroup%
  \renewcommand{\@caption@fignum@sep}{ (Color online). }%
  \caption[#1]{#2}%
  \endgroup%
}
\makeatother
\usepackage{amsmath}
\begin{document}
\title{Study of $\beta^+$/EC-decay properties of $sd$ shell nuclei using nuclear shell model}
\author{Surender$^1$, Vikas Kumar$^1$\thanks{vikasphysics@bhu.ac.in}, Praveen C.~Srivastava$^2$\thanks{praveen.srivastava@ph.iitr.ac.in}}
\institute{$^{1}$Department of Physics, Institute of Science, Banaras Hindu University Varanasi, Varanasi - 221 005, INDIA \\
$^{2}$Department of Physics, Indian Institute of Technology Roorkee, Roorkee 247 667, 
INDIA }

\date{\today}

%========================================================================
\abstract{Our study employs the nuclear shell model to systematically compute the half-lives of $\beta$ -decay for nuclei in the mass range of $A = 18-39$, encompassing the majority of $sd$ shell nuclei. This analysis utilizes the USDB and SDNN Hamiltonians. The theoretical outcomes contain calculations of various parameters such as $Q$ -values, half-lives, excitation energy, log$ft$ values, and branching ratios. We explore these results with axial-vector coupling constant for weak interactions, denoted as $g_A$$(= 1.27)$, and $\kappa$ value $(= 6289)$. 
We perform calculations of Gamow Teller matrix elements for 116 decay processes to calculate the quenching factor; we found a quenching factor of $q = 0.794\pm0.05 $ for the USDB interaction and $q = 0.815\pm0.04 $ for the SDNN interaction. We have also calculated superallowed transitions $0^+ \rightarrow  0^+$ for seven nuclei. Further, we have also included the electron capture phase space factor for the required nuclei to calculate the half-lives. This inclusion leads to small contribution in results, particularly for nuclei where electron capture (EC) plays a significant role.
The overall results are in agreement with the experimental data.}

\authorrunning{Vikas Kumar and P.C. Srivastava}
\maketitle

%\nonindent
%%%%%%%%%%%%%%%%%%%%%%%%%%%%%%%%%%%%%%%%%%%

\section{Introduction}
\label{introduction}

The $\beta^+$ and EC -decay is the dominant mode of radioactive decay for unstable proton-rich nuclei; in $\beta^+$ -decay process a proton changes into a neutron and in EC -decay process inner orbital electron of atom is absorbed within the nucleus where it combines with a proton and forming a neutron and a neutrino \cite{Gysbers}. The $\beta$ -decay studies are important because we can get information on both nuclear structure and nuclear astrophysics \cite{Algora,Rubio}. The nuclear shell model has successfully described various aspects of nuclear structure and beta decay properties of exotic nuclei \cite{Otsuka}.
The nuclear structure study of $N = Z$ nuclei in the nuclear chart is fascinating because these nuclei show several phenomena, such as shape coexistence along the $N = Z$ line and the role of pairing correlation of neutron and proton \cite{S. Akkoyun}. From the $rp$ -process point of view, $N = Z$ nuclei are also important. 
In this work, we have used the nuclear shell model approach to study a theoretical understanding of observed $\beta^+$-decay properties of 37 proton-rich and $N = Z$ nuclei of $sd$ space. 
The present study is motivated with the availability of recent experimental data for $\beta^+$-decay.

In Wilson {\it et al.} \cite{Wilson}, the experimental Gamow-Teller $\beta^+$ and electron-capture decays of $^{14}$O, $^{21}$Na, $^{25}$Al, $^{26}$Si, $^{29}$P, $^{30}$P, $^{30}$S, $^{31}$S, $^{33}$Cl, $^{34}$Cl, $^{35}$Ar, $^{38}$Ca, and $^{41}$Sc nuclei have reported and compared with previous works. They also reported thirteen new observed branching ratios for these nuclei. 
The $\beta^+$ -decays of $^{18}$Ne and $^{19}$Ne and their relation to parity mixing in $^{18}$F and $^{19}$F is reported in \cite{18ne}. The Gamow–Teller branching ratio and half-life in the $\beta$ decay of $^{21}$Na are outlined in \cite{Achouri2}. 
The observed half-life and branching ratio of $^{21}$Mg are reported in \cite{Sextro}.
Experimental investigations on the half-lives and superallowed $\beta$ decay of three nuclei, namely $^{18}$Ne, $^{22}$Mg, and $^{26}$Si, were studied in \cite{Hardy}. 
Using the IGISOL facility at the University of Jyväskylä, half-lives and branching ratios of $^{23}$Mg and $^{27}$Si have been measured and reported in \cite{Magron}.

In Achouri {\it et. al.} \cite{Achouri}, the half-life of $^{22}$Al was observed through $\beta^+$ decay process $^{22}$Al $\rightarrow$ $^{22}$Mg at GANIL using LISE3 facility. 
%The $\beta^-$delayed $\gamma$ and proton decay of $^{23}$Al were investigated in \cite{Saastamoinen} at the MARS facility at Texas A\&M University. 
In Warburton et al. \cite{Warburton} $\beta^+$-deacy of $^{24}$Al has been studied, and a few new branches are also reported.
The $\beta$ decay properties of the neutron-deficient nuclei $^{25}$Si and $^{26}$P were explored at GANIL using the LISE3 facility \cite{Thomas}.
Several weak branches are observed for $^{27}$Si and $^{35}$Ar $\beta^+$-decay in Daehnick et. al. \cite{Daehnick}. 
The $ft$ value of the superallowed $0^{+}$ $\rightarrow$ $0^{+}$ transition in $\beta^+$-decay of $^{32}$Ar has been discussed in \cite{Bhattacharya}. The proton-rich nucleus $^{33}$Ar has been studied through $\beta^+$-decay using the SPIRAL facility at GANIL. In Iacob et. al. \cite{Iacob}, the $\beta^+$-decay half-lives of $^{34}$Ar and $^{34}$Cl are measured and discussed in details. 
The $\beta^+$ decay half-life of neutron-deficient isotope $^{31}$Cl was reported in \cite{Bennett}. 
First time the $^{35}$K nucleus has been populated and its decay properties have been studied in \cite{Ewan}.
The $\beta$ decay of $^{37}$K is studied in \cite{Hagberg}, and several new $\beta$ decay branches are assigned to this nucleus. 
%New transitions in $\beta^-$ decay of $^{36}$Ca were studied using the LISE3 facility at GANIL in \cite{Lopez Jimenez}. 
In Kaloskamis et. al. \cite{Kaloskamis}, the $\beta^+$-decay properties of $^{37}$Ca are reported. The $\beta^+$- decay of $^{38}$Ca and its branching ratios are observed using the LISE3 facility of GANIL in \cite{Blank}. The $\beta^+$ decays of $^{39}$Ca and $^{35}$Ar were measured by using a Ge(Li) detector in \cite{Adelberger}.

In the present work first we have calculated the quenching factor for ${sd}$ space using Gamow-Teller transition strength of ${\beta^+}$-decay and  then we did a comprehensive $\beta^+$/EC -decay shell model study of 37 ${sd}$ shell nuclei. Further, the calculated excitation energies, log${ft}$ -values, quenched half-lives, ${Q}$ -values and branching ratios are compared with the available experimental data. 
In our previous work we have reported the shell model study of ${\beta}$ -decay for ${fp}$ and ${fpg}$ shell nuclei in \cite{vikas1,vikas3,vikas4}, and ${\beta^+}$/EC-decay study for ${Z = 21-30}$ shell nuclei in \cite{vikas2}. 

The section 2 consists of $\beta^+$/EC-decay formalism, $Q$-value formalism, shell model Hamiltonian, model space, and quenching factor. The results and discussions are given in section 3. Finally, the summary and conclusions are drawn in section 4.

\section{Details of the Shell Model Calculations}

%\subsection{${\beta}$-decay formalism}
\subsection{\texorpdfstring{$\beta$}{beta}-decay formalism}
 
 The observed Gamow Teller strength appears to be systematically smaller than what is theoretically expected on the basis of the model-independent Ikeda sum rule ‘$3(N-Z)$’. The quench factor $q$ in a given model space is calculated by averaging all ratios between the experimental and theoretical matrix elements $R(GT)$ values. The $R(GT)$ values are calculated using Gamow–Teller reduced transition probabilities $B(GT)$. Following Ref. \cite{martinez}, the Gamow–Teller reduced transition probabilities is given by
\begin{equation}\label{eq:1}
 B(GT)= \left(\frac{g_A}{g_V}\right)^2{\langle{\sigma\tau}\rangle}^2,
\end{equation}
where the axial-vector coupling constant of the weak interactions is represented by $g_A$ (= 1.270) and $g_V$ (= 1.0) represents the vector coupling constant of the weak interaction. The Gamow-Teller operator ${\langle{\sigma\tau}\rangle}$ is given by 
\begin{equation}\label{eq:2}
 {\langle{\sigma\tau}\rangle} = \frac{{\langle {f}|| \sum_{k}{\sigma^k\tau_{\pm}^k} ||i \rangle}}{\sqrt{2J_i + 1}},
\end{equation}
where $i$ and $f$ represent the required quantum numbers to specify the initial and final states, respectively. The $\pm$ refers to $\beta^{\pm}$ decay, $\tau_{\pm} = \frac{1}{2}(\tau_x + i\tau_y)$ with $\tau_+p$ = $n$, $\tau_-n$ = $p$.
The total angular momentum of the initial state is represented by $J_i$.

The matrix elements $M(GT)$ are related to reduced transition probability $B(GT)$ as \cite{B. A. Brown1},
\begin{equation}\label{eq:3}
M(GT)= [(2J_{i}+1)B(GT)]^{1/2}.
\end{equation}
The ``expected" total strength W can be used to normalize the matrix elements $M(GT)$ to get effective $g_{A}$. The total strength W is defined as

\begin{equation}\label{eq:4}
  W=\left\{
  \begin{array}{@{}ll@{}}
    |g_{A}/g_{V}|[(2J_{i}+1)3|N_{i}-Z_{i}|]^{1/2} , & for N_{i} \neq  Z_{i},\\
    |g_{A}/g_{V}|[(2J_{f}+1)3|N_{f}-Z_{f}|]^{1/2} , & for N_{i} = Z_{i},
  \end{array}\right.
\end{equation}

The matrix elements $R(GT)$ are defined as
\begin{equation}\label{eq:5}
R(GT) = \frac{M(GT)}{W}.
\end{equation}

The experimental $B(GT)$ values are calculated using experimental log$ft$ values (listed in Table \ref{table1}), the experimental $M(GT)$ values are obtained using Eq. \ref{eq:3}.

In $\beta^+$/EC-decay process, the mass number $A$ remains unchanged, whereas the atomic number $Z$ changes by one unit. The half-life ($t_{1/2}$) of $\beta$-decay is related to transition probability ($T_{fi}$) as

\begin{equation}\label{eq:6}
 t_{1/2}=\frac{ln2}{T_{fi}}.
\end{equation}

The consequential expression for total decay half-life of a combined $\beta^+$ and EC transition (represented by $\beta^+$/EC) is given by
\begin{equation}\label{eq:7}
 f_{0}t_{1/2}=\left[f_0^{(+)}+f_0^{(EC)}\right]t_{1/2} = \frac{\kappa}{[g^2_A*B(GT) + B(F)]},
\end{equation}
where $f_0$ is the phase-space factor (Fermi integral). 
The $B(F)$ is the Fermi reduced transition probabilities. The updated value of $\kappa$ from literature \cite{Patrignani} is given by

\begin{equation}\label{eq:8}
 \kappa \equiv \frac{2\pi^3\hbar^7ln2}{m^5_ec^4{(G_Fcos{\theta}_C)}^2} = 6289s,
\end{equation}
where the ${\theta}_C$ is the Cabibbo angle. 

The Fermi reduced transition probability $B(F)$ is given by
\begin{equation}\label{eq:9}
 B(F) \equiv \frac{g^2_V}{{2J_i + 1}}{|M_F|^2},
\end{equation}
where $M_F$ is the Fermi matrix element.

The phase-space factor for $\beta^{+}$ -decay is given by 
 
\begin{equation}\label{eq:10}
 f_0^{(+)}= \int_{1}^{E_0} F_0({-}Z_f, \epsilon)p\epsilon(E_0-\epsilon)^2 \,d\epsilon,
\end{equation}
where $F_0({-}Z_f, \epsilon)$ is called the Fermi function and 
\begin{equation}\label{eq:11}
 \epsilon \equiv \frac{E_e}{m_ec^2}, E_0 \equiv \frac{E_i-E_f}{m_ec^2}, p \equiv \sqrt{\epsilon^2-1},
\end{equation}
where $E_e$ is the total energy of the emitted positron and $E_i$ and $E_f$ are the energies of the initial and final nuclear states, respectively.\\
%%%%%%%%%%%%%%%%%%%%%%%%%%%%%%%%%%%%%%%%%%%%%
%%%%%%%%%%%%%%%%%%%%%%%%%%%%%%%%%%%%%%%%%%%%%%%%%%%
    Following Ref. \cite{suhonen}, the phase-space factor for the EC transition is given by
\begin{equation}\label{eq:12}
f^{(EC)}_0 = 2\pi({\alpha}Z_i)^3(\epsilon_0+E_0)^2,
\end{equation}
where $\epsilon_0$ is given by
\begin{equation}\label{eq:13}
\epsilon_0 \equiv 1-\frac{1}{2}({\alpha}Z_i)^2,
\end{equation}
and $\alpha$ $(=\frac{1}{137})$ represents the fine-structure constant. The non-relativistic $s$-electron wave function are considered in Eq. \ref{eq:13}, and this approximation not holds good for small decay energies, additional corrections arise from the screening of the nuclear charge by atomic electrons and from the finite nuclear size.

The experimental $\beta$ -decay $Q$ values are taken from \cite{ENSDF}.
%}
%%%%%%%%%%%%%%%%%%%%%%%%%%%%%%%%%%%%%%%%%%%%%%%%%%%%%
%%%%%%%%%%%%%%%%%%%%%%%%%%%%%%%%%%%%%%%%%%%%%%%%%%%%%%%%%

Usually, $ft$ values are large, expressed in terms of `log $ft$ values'. The log$ft$ $\equiv$ log$_{10}({f_0}t_{1/2}[s])$.

  The total half-life can be calculated with the help of the partial half-life ($t_{i}$) of the daughter states $i$ using the following expression: 
\begin{equation}\label{eq:14}
 \frac{1}{t_{1/2}}= \left({\sum_i {\frac{1}{t_i}}}\right).
\end{equation}

The expression for the partial half-life of the allowed $\beta$-decay is taken from \cite{suhonen}.

The branching ratio ${b_r}$ is related to partial half-life ${t_i}$ and the total half-life $t_{1/2}$ of the allowed $\beta$-decay as
\begin{equation}\label{eq:15}
%\label{br}
 t_{i}= {\frac{t_{1/2}}{b_r}}.
% \label{br}
\end{equation}
where $t_i$ is the partial half-life and $t_{1/2}$ is the total half-life of the allowed $\beta$-decay.

\subsection{The \texorpdfstring{$Q$}{Q}-value formalism}

The theoretical $\beta^+$-decay $Q$-values were obtained using following relation \cite{sweta}
%\begin{equation}
%\label{eq:20}
%\label{qvalue}
%Q = \left[E(SM)_i + E(C)_i\right] - \left[E(SM)_f + E(C)_f\right],
%\end{equation}
\begin{equation}
\label{eq:16}
%\label{qvalue}
Q(\beta^{+}) = E^{parent}_{g.s}-E^{daughter}_{g.s} + \delta{m},
\end{equation}
where $E^{parent}_{g.s}$ and $E^{daughter}_{g.s}$ represent the ground state binding energies of parent and daughter nuclei, respectively, and $\delta{m}=(m_p-m_n-m_{e^{+}})c^2=-1.802$ MeV.  

The binding energy of the ground state is given by
\begin{equation}
\label{eq:17}
%\label{qvalue}
E = E_{SM} + E_{core} + E_{C},
\end{equation}
where $E_{SM}$ is the calculated binding energy using shell-model and
$E_{core}$ is the binding energy of the considered core in SM calculations, and
$E_{C}$ is the valence space Coulomb energy.
In \cite{Caurier}, the expression for $E_{C}$ is given as
\begin{equation}\label{eq:18}
E_{C} = 0.700[Z(Z-1)-0.76(Z(Z-1))^{2/3}]/R_C,
\end{equation}
\begin{equation}
R_C = e^{1.5/A}A^{1/3}\left(0.946-0.573\left( \frac{2T}{A}\right)^2\right),
\end{equation}
where $T$ is the isospin of the nucleus.
\subsection{{Shell Model Hamiltonian, Model space and quenching factor}}
The $\beta^+$-decay half-lives calculation has been performed for nuclei in $sd$ model space using USDB \cite{brown} and SDNN \cite{S. Akkoyun} effective interactions. 
The NuShellX@MSU \cite{MSU-NSCL} shell model code has been used to diagonalize energy matrices. The USDB Hamiltonian is the mass-dependent version of USD \cite{Wildenthal,B. A. Brown1} and the mass dependence of TBMEs have the form $[18/(16+n)]^{0.3}$, where the number of valence nucleons is represented by $n$. The single-particle energies for USDB Hamiltonian are taken to be 2.1117, -3.9257, and -3.2079 in MeV for the $d_{3/2}$, $d_{5/2}$ and $s_{1/2}$ orbits, respectively. The USDB interaction is characterized by 66 parametric values, viz., three bare single-particle energies and 33 and 30 TBMEs of T = 0 and 1, respectively. 
In the SDNN interaction, Hamiltonian  \cite{S. Akkoyun}, the USDB interaction was re-estimated for $N = Z$ nuclei using artificial neural networks. 
  
\begin{figure*}
\begin{center}
\resizebox{0.98\textwidth}{0.40\textwidth}{\includegraphics{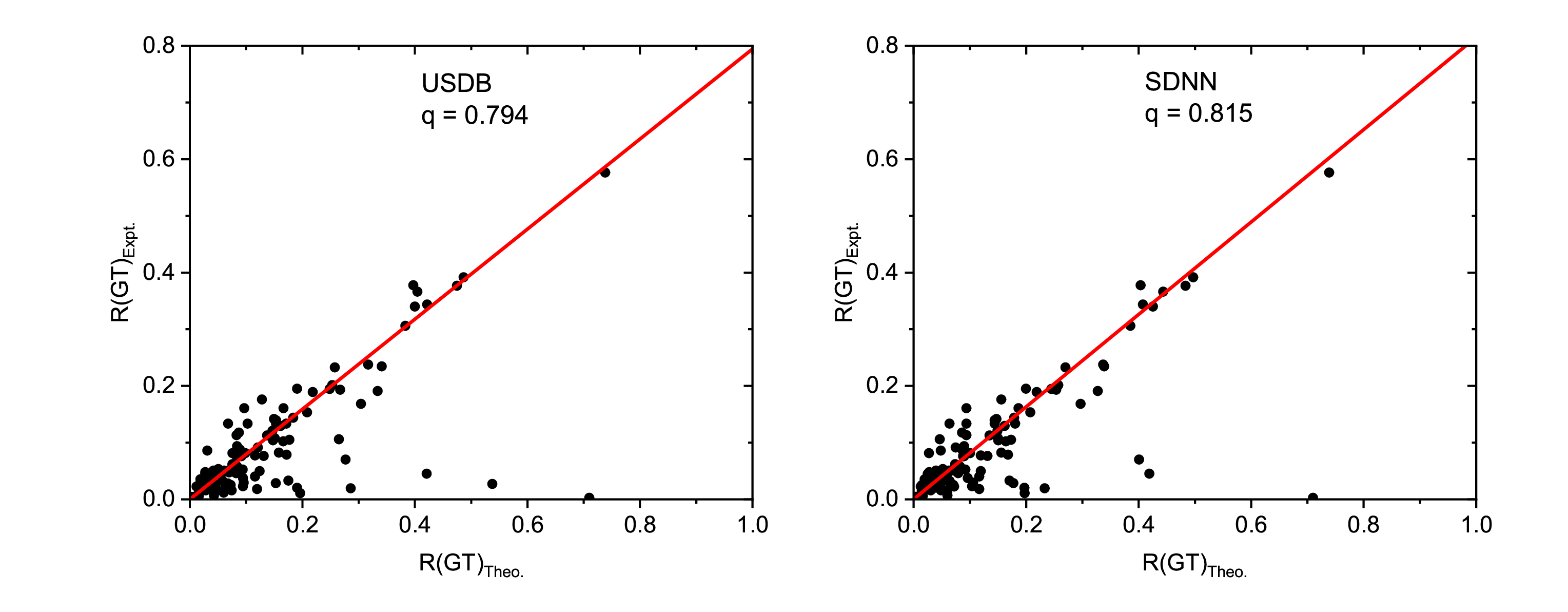}}
\caption{\label{fig1} The experimental versus theoretical matrix element $R(GT)$ values. SM calculations are based on the ``free-nucleon" Gamow-Teller operator. Each transitions are represented by a point in the graph. The theoretical and experimental values are taken on the x and y axes, respectively.}
\end{center}
\end{figure*}

The theoretical SM results of Gamow-Teller transition strengths are more significant than the experimental ones. Hence, we need a quenching factor (q) for this model space to better agreement between theoretical and experimental results. Using Eq. \ref{eq:3} - \ref{eq:5}, we have calculated theoretical and experimental $R(GT)$ values for a particular model space by averaging all the ratios between the experimental and theoretical $R(GT)$ values we get quenching factor (q). The comparison between experimental and theoretical $R(GT)$ values are plotted in Fig.~\ref{fig1}. To get experimental $R(GT)$ values, we took experimental log$ft$ values, which are listed in Table \ref{table1}. In Fig.~\ref{fig1}, the slope of the straight line gives the average quenching factor. In the present work, we have obtained two different quenching factors for $sd$ shell nuclei, $q$ = 0.794$\pm$0.05 using USDB, and  $q$ = 0.815$\pm$0.04 using SDNN effective interactions.

%%%%%%%%%%%%%%%%%%%%%%%%%%%%%%%%%%%%%%%%%%%%%%%%%%%%%%%%%%%%%%%%%%%%%%%%%%%%%%%%%%%%%%%%%%%%%%%%%%%%%%%%%%%%
\begin{table*}
  \begin{center}
    \leavevmode
    \caption{\label{table1}Listed the experimental and theoretical $M(GT)$ matrix elements; $I_{\beta}+I{\epsilon}$ are the branching ratios; $J^{\pi}_{n}$ and $T^{\pi}_{n}$ are the spin-parity and isospin of the final states, respectively.}

\begin{tabular}{lccccccccccccc}
\hline
&  &  & $I_{\beta}+I{\epsilon}$ &  &\multicolumn{3}{c}{$M(GT)$}&&  \\
\cline{6-8}
Process &$2J^{\pi}_{n}$, $2T^{\pi}_{n}$& $Q$(MeV) &$(\%)$& log$ft$ & \multicolumn{1}{c}{EXPT.}&\multicolumn{1}{c}{USDB}&\multicolumn{1}{c}{SDNN}&$W$\\
\cline{3-4}
\hline
$^{18}$Ne$(\beta^+)$$^{18}$F&$2_1^+,0$&4445.7(47)&92.08&3.091&1.778&2.279&2.279&3.111\\
                            &$2_2^+,0$&2744.8&0.188&4.47&0.363&0.269&0.266&\\
%^{19}$Ne$(\beta^+)$$^{19}$F&$1_1^+,1$&3238.4&99.88&3.2329&0.026&0.071&0.068&3.111\\
$^{20}$Na$(\beta^+)$$^{20}$Ne&$6_1^+,0$&4403&0.028&5.77&0.182&0.436&0.454&6.956\\
                             &$6_2^+,0$&3002&0.117&4.36&0.923&0.709&0.649&\\
$^{21}$Na$(\beta^+)$$^{21}$Ne&$5_1^+,1$&3196.4&5.067&4.596&0.629&0.802&0.779&4.399\\
                             &$1_1^+,1$&753&--&4.61&0.619&0.653&0.643&\\
$^{22}$Na$(\beta^+)$$^{22}$Ne&$4_1^+,2$&1568.67&99.94&7.41&0.033&0.053&0.065&6.956\\
$^{21}$Mg$(\beta^+)$$^{21}$Na&$3_1^+,1$&13098(16)&16&5.26&0.359&0.295&0.292&9.333\\
                             &$7_2^+,1$&11382&10.9&5.11&0.426&0.761&0.736&\\
$^{22}$Mg$(\beta^+)$$^{22}$Na&$2_1^+,0$&4198.55&41.32&3.64&0.945&1.182&1.189&3.111\\
                             &$2_2^+,0$&2844.7&5.448&3.46&1.163&1.466&1.491&\\
$^{23}$Mg$(\beta^+)$$^{23}$Na&$5_1^+,1$&3615.97&7.849&4.434&0.758&0.966&0.952&4.399\\
                             &$1_1^+,1$&1665.27&0.006&4.97&0.409&0.689&0.673&\\
$^{22}$Al$(\beta^+)$$^{22}$Mg&$6_1^+,2$&13149.2&5.9&5.56&0.311&0.538&0.520&13.198\\
                             &$6_2^+,2$&11736&3&5.6&0.297&0.153&0.170&\\
$^{22}$Al$(\beta^+)$$^{22}$Mg&$10_1^+,2$&11469&18.5&4.75&0.790&0.991&0.979&\\
$^{23}$Al$(\beta^+)$$^{23}$Mg&$3_1^+,1$&12221.6(4)&36.3&5.3&0.342&0.281&0.267&9.333\\
$^{25}$Al$(\beta^+)$$^{25}$Mg&$3_2^+,1$&3301.95&0.047&6.2&0.121&0.173&0.182&5.388\\
                             &$7_1^+,1$&2664.93&0.783&4.36&1.011&1.167&1.170&\\
$^{25}$Si$(\beta^+)$$^{25}$Al&$3_1^+,1$&11795&26&5.05&0.457&0.563&0.580&9.333\\
                             &$3_2^+,1$&10068&4.8&5.42&0.298&0.542&0.593&\\
                             &$3_3^+,1$&8551&2.99&5.25&0.363&0.368&0.357&\\
                             &$3_4^+,1$&6570&0.32&5.6&0.242&0.416&0.412&\\
                             &$3_5^+,1$&5633&3.7&4.16&1.272&0.357&0.357&\\
$^{26}$Si$(\beta^+)$$^{26}$Al&$2_1^+,0$&4011.4&21.92&3.54&1.060&1.302&1.258&3.111\\
                             &$2_2^+,0$&3218.52&2.72&3.86&0.734&0.978&1.040&\\
                             &$2_3^+,0$&1345.33& $\le$$5\cdot10^{-4}$&4.2&0.496&0.514&0.576&\\
$^{27}$Si$(\beta^+)$$^{27}$Al&$3_1^+,1$&3797.82&0.006&7.23&0.037&0.228&0.322&5.388\\
                             &$7_1^+,1$&2600.25&0.179&4.69&0.691&0.864&0.866&\\
                             &$3_2^+,1$&1830.18&0.025&4.34&1.034&1.427&1.355&\\
$^{26}$P$(\beta^+)$$^{26}$Si&$4_1^+,2$&16460.73&44&4.9&0.586&0.854&0.835&11.640\\
                             &$4_2^+,2$&15741.69&3.3&5.9&0.185&0.099&0.075&\\
                             &$8_1^+,2$&14415.8&1.68&6.0&0.165&0.187&0.169&\\
                             &$4_3^+,2$&14118.82&1.78&5.9&0.294&0.442&0.429&\\
                             &$4_4^+,2$&11963&0.78&5.9&0.415&0.037&0.026&\\
$^{28}$P$(\beta^+)$$^{28}$Si&$4_1^+,0$&12556.07&69.1&4.851&0.620&0.750&0.720&8.231\\
                             &$8_1^+,0$&9727.57&1.29&5.99&0.167&0.298&0.305&\\
$^{29}$P$(\beta^+)$$^{29}$Si&$3_1^+,1$&3669.11&1.263&4.806&0.349&0.256&0.291&3.111\\
                             &$3_2^+,1$&2516.52&0.45&4.172&0.724&1.053&1.046&\\
$^{30}$P$(\beta^+)$$^{30}$Si&$0_1^+,2$&4232.4(3)&99.94&4.839&0.412&0.473&0.444&6.956\\
                            &$4_1^+,2$&1997.08&0.052&5.884&0.124&0.169&0.210&\\
                            &$4_2^+,2$&733.85&0.002&5.26&0.254&0.372&0.331&\\
                            &$0_2^+,2$&462.88&0.003&4.48&0.622&0.781&0.794&3.111\\
$^{29}$S$(\beta^+)$$^{29}$P&$3_1^+,1$&12411.45&27&5.06&0.451&0.735&0.738&9.333\\
                           &$3_2^+,1$&11372.3&20.7&4.98&0.495&0.467&0.481&\\
$^{29}$S$(\beta^+)$$^{29}$P&$7_1^+,1$&9714.5&3.9&5.03&0.467&0.643&0.713&\\
$^{30}$S$(\beta^+)$$^{30}$P&$2_1^+,0$&6138(3)&21.3&4.322&0.431&0.472&0.444&3.111\\
                           &$2_2^+,0$&5429.3&0.29&5.89&0.071&0.291&0.321&\\
                           &$2_3^+,0$&3118.8&2.283&3.549&1.050&1.235&1.313&\\
%$^{31}$S$(\beta^+)$$^{31}$P&$1_1^+,1$&5398.01&98.79&3.679&1.279&0.759&0.737&3.111\\
%                           &$1_1^+,1$&2263.71&0.0317&4.76&0.368&0.387&0.407&\\
$^{31}$Cl$(\beta^+)$$^{31}$S&$1_1^+,1$&12008(3)&~7.0&5.6&0.198&0.274&0.279&7.620\\
                           &$5_1^+,1$&9773.94&38&4.37&0.816&1.139&1.133&\\
\hline
\end{tabular}
%\end{small}
\end{center}
\end{table*}

\addtocounter{table}{-1}

\begin{table*}
  \begin{center}
    \leavevmode
\caption{{\em Continuation.\/}}

\begin{tabular}{lrcccccccccccc}
\hline
&  &  & $I_{\beta}+I{\epsilon}$ &  &\multicolumn{3}{c}{$M(GT)$}&&  \\
\cline{6-8}
Process &$2J^{\pi}_{n}$, $2T^{\pi}_{n}$& $Q$(MeV) &$(\%)$& log$ft$ & \multicolumn{1}{c}{EXPT.}&\multicolumn{1}{c}{USDB}&\multicolumn{1}{c}{SDNN}&$W$\\
\cline{3-4}
\hline
                           &$1_2^+,1$&8931.56&2.54&5.33&0.270&0.289&0.258&\\
                           &$5_2^+,1$&8724.24&4.46&5.03&0.382&0.487&0.485&\\
$^{33}$Cl$(\beta^+)$$^{33}$S&$1_1^+,1$&4741.69&0.48&5.65&0.187&0.229&0.238&4.399\\
                            &$5_1^+,1$&3615.99&0.461&4.97&0.409&0.367&0.392&\\
                            &$5_2^+,1$&2715.09&0.436&4.18&1.015&1.125&1.179&\\
                            &$5_3^+,1$&1750.29&$\le$ $7\cdot10^{-4}$&6.4&0.079&0.521&0.509&\\
%&$5_2^+,1$&2715.09&0.436&4.180&1.015&1.125&1.179&4.399\\
                            &$1_2^+,1$&1529.39&$1\cdot10^{-4}$&6.2&0.099&0.312&0.314&\\
                            &$1_3^+,1$&1207.09&$\le$$4\cdot10^{-4}$&5.7&0.176&0.505&0.510&\\
$^{34}$Cl$(\beta^+)$$^{34}$S&$4_1^+,2$&3364.07&28.52&5.982&0.169&0.242&0.196&6.956\\
                            &$4_2^+,2$&2187.42&26.42&4.823&0.641&0.910&0.926&\\
                            &$4_3^+,2$&1376.82&0.457&5.081&0.476&0.711&0.736&\\
$^{34}$Cl$(\beta^+)$$^{34}$S&$8_1^+,2$&802.65&0.033&5.44&0.315&0.286&0.306&9.333\\
$^{32}$Ar$(\beta^+)$$^{32}$Cl&$2_1^+,2$&11134.7&$\le$7&$\ge$5.1&0.155&0.082&0.085&4.399\\
                            &$2_2^+,2$&9966.15&57&3.94&0.669&0.909&0.907&\\
                            &$2_3^+,2$&8924.5&0.35&5.9&0.070&0.215&0.214&\\
                            &$2_4^+,2$&7362.7&3.681&4.424&0.383&0.388&0.376&\\
                            &$2_5^+,2$&6967.7&0.9&4.9&0.222&0.184&0.174&\\
                            &$2_6^+,2$&6695.7&0.117&5.7&0.088&0.832&0.859&\\
                            &$2_7^+,2$&6338.7&0.053&5.91&0.069&0.324&0.237&\\
                            &$2_8^+,2$&5832.7&0.115&5.371&0.129&0.417&0.461&\\
                            &$2_9^+,2$&5439.7&0.231&4.9&0.222&0.371&0.360&\\
                            &$2_{10}^+,2$&5071.7&0.146&4.92&0.217&0.541&0.521&\\
                            &$2_{11}^+,2$&4606.7&0.049&5.16&0.164&0.412&0.420&\\
                            &$2_{12}^+,2$&4396.7&0.127&4.62&0.306&1.207&1.750&\\
                            &$2_{13}^+,2$&3804.7&0.125&4.26&0.463&1.157&0.202&\\
                            &$2_{14}^+,2$&3526.7&0.007&5.4&0.125&0.666&0.774&\\
                            &$2_{15}^+,2$&3273.7&0.046&4.49&0.355&0.330&0.119&\\
$^{33}$Ar$(\beta^+)$$^{33}$Cl&$3_1^+,1$&11619.1(6)&18.7&5.022&0.272&0.324&0.335&5.388\\
                            &$3_2^+,1$&9266.8&1.7&5.54&0.150&0.361&0.351&\\
                            &$3_3^+,1$&7648.1&0.382&5.744&0.119&0.166&0.183&\\
                            &$3_4^+,1$&7506.1&0.453&5.626&0.136&0.388&0.378&\\
$^{34}$Ar$(\beta^+)$$^{34}$Cl&$2_1^+,0$&5601.03&0.91&5.31&0.138&0.084&0.077&3.111\\
                            &$2_2^+,0$&5396.03&2.492&4.78&0.254&0.489&0.481&\\
                            &$2_3^+,0$&3482.33&0.864&4.12&0.544&0.396&0.482&\\
                            &$2_4^+,0$&2932.63&1.29&3.458&1.165&1.226&1.245&\\
$^{35}$Ar$(\beta^+)$$^{35}$Cl&$1_1^+,1$&4746.88&1.232&5.091&0.356&0.430&0.438&4.399\\
                            &$5_1^+,1$&4202.99&0.248&5.475&0.229&0.410&0.403&\\
                            &$5_2^+,1$&2963.8&0.088&4.973&0.400&0.528&0.326&\\
                            &$1_2^+,1$&1998.4&0.007&4.81&0.491&0.599&0.589&\\
$^{35}$K$(\beta^+)$$^{35}$Ar&$5_1^+,1$&10123.8&11.901&4.91&0.438&0.636&0.623&7.620\\
                            &$5_2^+,1$&8891.7&26&4.27&0.915&1.111&1.129&\\
$^{35}$K$(\beta^+)$$^{35}$Ar&$1_1^+,1$&7168.6&2.1&4.85&0.469&0.574&0.559&\\
$^{37}$K$(\beta^+)$$^{37}$Ar&$1_1^+,1$&4737.66&0.004&7.38&0.025&0.069&0.072&4.399\\
                            &$5_1^+,1$&3351.38&2.067&3.785&1.600&1.765&1.936&\\
                            &$5_2^+,1$&2977.68&0.003&6.34&0.084&1.248&1.017&\\
$^{38}$K$(\beta^+)$$^{38}$Ar&$4_1^+,2$&3746.47&99.84&4.975&0.538&0.922&0.922&6.956\\
                            &$4_2^+,2$&1978.27&0.151&5.88&0.190&3.771&0.154&\\
$^{36}$Ca$(\beta^+)$$^{36}$K&$2_1^+,2$&9853.6&14.302&4.519&0.344&0.751&0.733&4.399\\
                            &$2_2^+,2$&9347.4&31.007&4.06&0.583&0.748&0.789&\\
                            &$2_3^+,2$&7609&10.304&4.06&0.583&0.297&0.278&\\   
                            &$2_4^+,2$&6516&2.602&4.29&0.447&0.725&0.717&\\
                            &$2_5^+,2$&6308&1.2&4.55&0.332&0.363&0.393&\\
\hline
\end{tabular}
\end{center}
\end{table*}

\addtocounter{table}{-1}

\begin{table*}
  \begin{center}
    \leavevmode
\caption{{\em Continuation.\/}}

\begin{tabular}{lrcccccccccccc}
\hline
&  &  & $I_{\beta}+I{\epsilon}$ &  &\multicolumn{3}{c}{$M(GT)$}&&  \\
\cline{6-8}
Process &$2J^{\pi}_{n}$, $2T^{\pi}_{n}$& $Q$(MeV) &$(\%)$& log$ft$ & \multicolumn{1}{c}{EXPT.}&\multicolumn{1}{c}{USDB}&\multicolumn{1}{c}{SDNN}&$W$\\
\cline{3-4}
\hline
                            &$2_6^+,2$&5040&2.204&3.73&0.852&0.831&0.871&\\
                            &$2_7^+,2$&4179&0.502&3.9&0.701&0.421&0.410&\\
$^{37}$Ca$(\beta^+)$$^{37}$K&$2_1^+,1$&10293.6&2.14&5.69&0.089&0.453&0.452&7.620\\
                            &$5_1^+,1$&8914&7.9&4.79&0.251&1.325&1.290&\\
                            &$5_2^+,1$&8425&4.5&4.9&0.222&0.201&0.370&\\
$^{38}$Ca$(\beta^+)$$^{38}$K&$2_1^+,0$&6283.7&2.842&4.8&0.249&0.277&0.276&3.111\\
                            &$2_2^+,0$&5044.5&19.478&3.426&1.209&1.502&1.534&\\
                            &$2_3^+,0$&3401.1&0.343&4.16&0.519&0.938&0.916&\\
                            &$2_4^+,0$&2886.46&0.025&4.8&0.249&0.010&0.010&\\
                            &$2_5^+,0$&2765.9&0.112&4.05&0.590&1.030&1.011&\\
                            &$2_6^+,0$&2567.3&0.004&5.3&0.140&1.299&1.293&\\
$^{39}$Ca$(\beta^+)$$^{39}$K&$1_1^+,1$&4002.16&0.0025&7.020&0.019&0.006&0.020&4.399\\
                            
\hline
\end{tabular}
\end{center}
\end{table*}

%%%%%%%%%%%%%%%%%%%%%%%%%%%%%%%%%%%%%%%%%%%%%%%%%%%%%%%%%%%%%%%%%%%%%%%%%%%%%%%%%%%%%%%%%%%%%%%%%%%%%%%%%%%%%%%%%%%%%%
\begin{figure*}
\begin{center}
\resizebox{1.0\textwidth}{1.2\textwidth}{\includegraphics{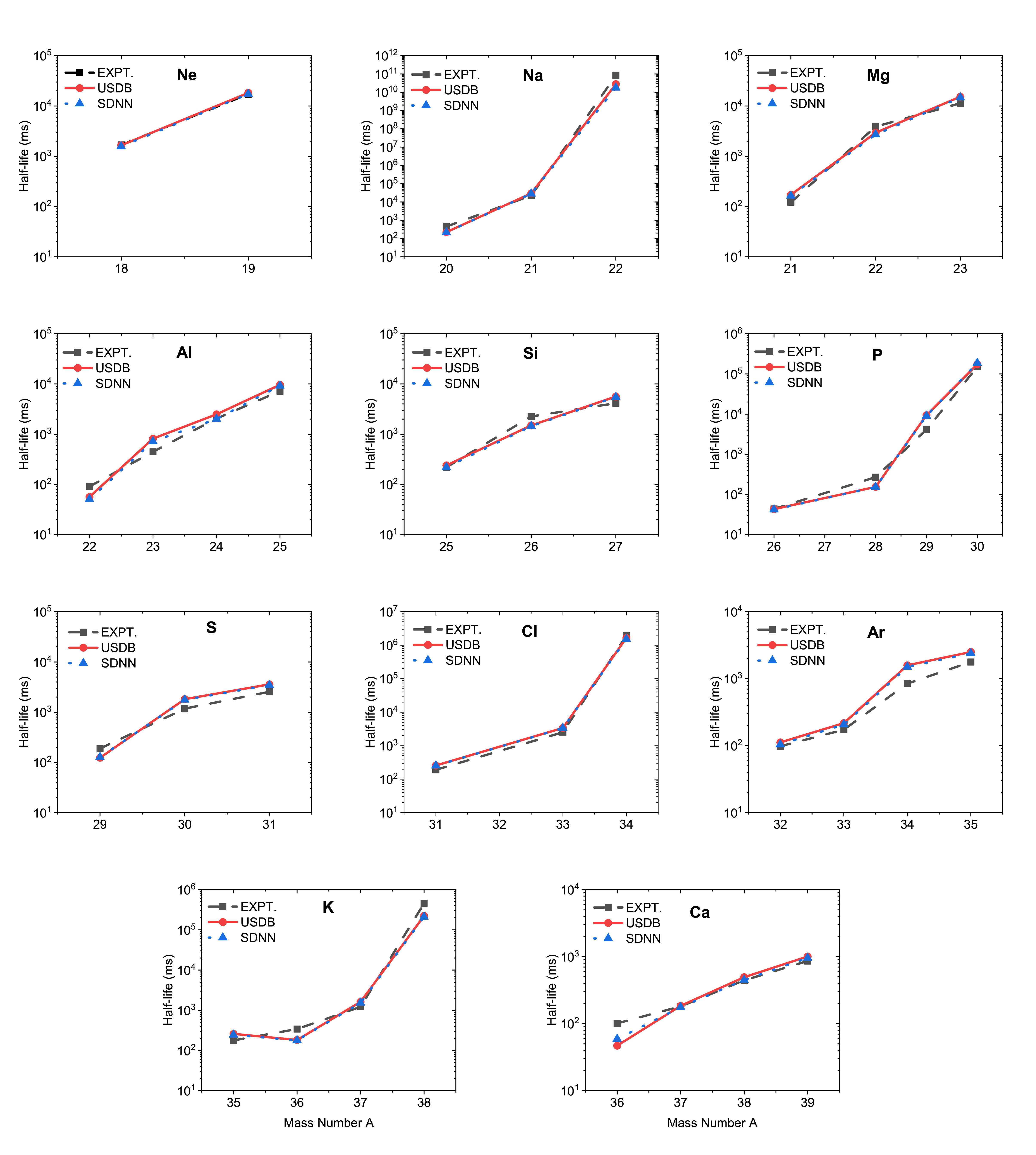}}
\caption{\label{fig2} The theoretical and experimental $\beta^+$/EC- decay half-life versus mass number A of the concerned $sd$ shell nuclei.}
\end{center}
\end{figure*}
%%%%%%%%%%%%%%%%%%%%%%%%%%%%%%%%%%%%%%%%%%%%%%%%%%%%%%%%%%%%%%%%%%%%%%%%%%%%
%%%%%%%%%%%%%%%%%%%%%%%%%%%%%%%%%%%%%
\begin{table*}
\begin{center}
\caption{\label{table2}Listed the theoretical phase space factors for $\beta^+$/EC decay (considered only those nuclei in which the EC phase space factors are significant); log$(f^{(+)}_0+f^{(EC)}_{0})$ values; the experimental $Q$ values taken from \cite{ENSDF}.}
\begin{tabular}{cccccc}
%\begin{tabular}{c@{\hspace{2pt}}c@{\hspace{2pt}}c@{\hspace{4pt}}c@{\hspace{4pt}}c@{\hspace{4pt}} c@{\hspace{4pt}}}
\hline
%\vspace{5pt}}
$^{A}Z_{i}(J^{\pi})$ &  $^{A}Z_{f}(J^{\pi})$ & $Q^{(EXPT.)}$ (MeV) & $f^{(+)}_0$ & $f^{(EC)}_0$ &log$(f^{(+)}_0+f^{(EC)}_{0})$\\
\hline
%\vspace{5pt}}
$^{22}$Na($3^+$)    & $^{22}$Ne($2_1^+$)    & 1.568  &   0.2806  &  0.0306   & -0.5072\\
%%%%%%%%%%%%%%%%%%%%%%%%%%%%%%%%%%%%%%%%%%%%%%%%%%%%%%%%%%%%%%%%%%%%%%%%%%%%%   
$^{23}$Mg($3/2^+$)  & $^{23}$Na($3/2_1^+$)   & 4.056  &  376.172    &  0.2658   & 2.5757  \\
                    & $^{23}$Na($5/2_1^+$)   & 3.615  &  186.951    &  0.2111   & 2.272  \\
                    & $^{23}$Na($1/2_1^+$)   & 1.665  &  0.5233    &  0.0497   & -0.2418  \\
%%%%%%%%%%%%%%%%%%%%%%%%%%%%%%%%%%%%%%%%%%%%%%%%%%%%%%%%%%%%%%%%%%%%%%%%%%%%%%%%%%%%%%%%%%%%%%%%%%%%%%%%%%%%%%%%%%%%%%%%%%%%%%%%%%% 
$^{29}$P($1/2^+$)   &$^{29}$Si($1/2_1^+$)    &  4.942 &  1125.19      & 0.7704  & 3.0515 \\
                    &$^{29}$Si($3/2_1^+$)    &  3.669 &  192.259   &  0.4244   & 2.2848\\
%                    &$^{29}$Si($5/2_1^+$)    & 2.914  &  44.0319    &  0.2676   & 1.6464\\
                    &$^{29}$Si($3/2_2^+$)    & 2.516  &  15.9367    &  0.1994   & 1.2078 \\
 
%%%%%%%%%%%%%%%%%%%%%%%%%%%%%%%%%%%%%%%%%%%%%%%%%%%%%%%%%%%%%%%%%%%%%%%%%%%%%%%%%%%%%%%%%%%%%%%%%%%%%%%%%%%%%%%%%%
$^{30}$P($1^+$)     &$^{30}$Si($0_1^+$)    & 4.232 &  455.214    & 0.5648    & 2.6587 \\
                    &$^{30}$Si($2_1^+$)    & 1.997 &  2.642    &  0.1256   &  0.4421\\
%%%%%%%%%%%%%%%%%%%%%%%%%%%%%%%%%%%%%%%%%%%%%%%%%%%%%%%%%%%%%%%%%%%%%%%%%%%%%%%%%%%%%%%%%%%%%%%%%%%%% 
$^{34}$Cl($3^+$)     &$^{34}$S($2_1^+$)    & 3.364 &  107.342    & 0.5191    & 2.0328 \\
                     &$^{34}$S($2_2^+$)    & 2.187 &  5.2964    &  0.2191   &  0.7416\\
                     &$^{34}$S($2_3^+$)    & 1.376 &  0.0396    &  0.0865   &  -0.8993\\
%%%%%%%%%%%%%%%%%%%%%%%%%%%%%%%%%%%%%%%%%%%%%%%%%%%%%%%%%%%%%%%%%%%%%%%%%%%%%%%%%%%%%%%%%%%%%%%%%%%%%%%%%%%%%%%%%%%%%%%%%%%%%%%%%%
$^{38}$K($3^+$)  &$^{38}$Ar($2_1^+$)        &  3.746 &  201.575   &   0.8983  & 2.3064 \\
%                 &$^{38}$Ar($3_1^+$)        &  2.103 &  3.6843  &  0.2825  & 0.5984 \\
                 &$^{38}$Ar($2_2^+$)        &  1.978 &  2.1988   &   0.2499  & 0.3889 \\
                 &$^{38}$Ar($2_3^+$)        &  1.348 &  2.6119   &   0.1158  & 0.4358 \\

%%%%%%%%%%%%%%%%%%%%%%%%%%%%%%%%%%%%%%%%%%%%%%%%%%%%%%%%%%%%%%%%%%%%%%%%%%%%%%%%%%%%%%%%%%%%%%%%%%%%%%%%%%%%%%%%%%%%%%%%%%%%%%%%%%
               
\hline 
\end{tabular}
%\label{be2}
\end{center}
\end{table*}
%%%%%%%%%%%%%%%%%%%%%%%%%%%%%%%%%%%%%
%%%%%%%%%%%%%%%%%%%%%%%%%%%%%%%%%%%%%%%%%%%%%%%%%%%%%%%%%%%
\pagestyle{empty}
\begin{table*}[htbp]
\begin{center}
 \begin{footnotesize}  
\caption{\label{table3}The theoretical log$ft$ values, excitation energies, and branching
percentages of $\beta^+$/EC -decays of the concerned nuclei are compared with the experimental data; References of the experimental data are given in the last column.}

\begin{tabular}{r@{\hspace{2pt}}c@{\hspace{0.1pt}}c@{\hspace{0.1pt}}c@{\hspace{0.1pt}}c@{\hspace{0.1pt}} c@{\hspace{0.1pt}}c@{\hspace{0.1pt}}c@{\hspace{0.1pt}}c@{\hspace{0.1pt}}c@{\hspace{0.1pt}}c@{\hspace{0.1pt}} c@{\hspace{0.5pt}}c@{\hspace{0.1pt}}}
\hline
&    &\multicolumn {3} {c} {Ex. energy (keV)} &\multicolumn {3} {c} {log$ft$ value}& \multicolumn {3} {c} {Branching ($\%$)}& Ref.\\
%\cline{3-5}
%\cline{6-8}
%\cline{9-11}
$^{A}$Z$_{i}(J^{\pi})$  & $^{A}$Z$_{f}(J^{\pi})$  &
   \multicolumn{1}{c}{USDB}& \multicolumn{1}{c}{SDNN}&\multicolumn{1}{c}{EXPT.} &
      \multicolumn{1}{c}{USDB}& \multicolumn{1}{c}{SDNN}&\multicolumn{1}{c}{EXPT.} &
      \multicolumn{1}{c}{USDB}& \multicolumn{1}{c}{SDNN}&\multicolumn{1}{c}{EXPT.}&
      \\
\hline
$^{18}$Ne(${0}^+$)&$^{18}$F($1^+$)&0&0&0&3.076&3.053&3.091&95.05&95.05&92.1&\cite{18ne}&\\
$^{19}$Ne(${1/2}^+$)&$^{19}$F(${1/2}^+$)&0&0&0&3.218&3.201&3.232&99.99&99.99&99.88&\cite{19ne}&\\
&$^{19}$F(${3/2}^+$)&1.770&1.789&1.554&7.791&7.769&5.71&$4\cdot10^{-6}$&$4\cdot10^{-6}$&0.002&&\\
$^{20}$Na(${2}^+$)&$^{20}$Ne($2^+$)&1.747&1.810&1.633&5.018&5.008&4.99&70.31&70.04&79.44&\cite{ENSDF}&\\
&&7.543&7.542&7.421&4.310&4.294&4.2&20.71&21.86&15.96&&\\
&&9.958&10.110&7.833&4.704&4.683&5.48&1.19&1.11&0.58&&\\
&&10.493&10.638&9.483&4.174&4.217&5.08&2.31&1.86&0.24&&\\
&&10.787&10.953&10.274&3.972&3.921&3.48&2.63&2.55&2.87&&\\
&&12.273&12.315&10.941&5.800&5.551&5.3&0.004&0.007&0.01&&\\
&&13.267&13.322&11.116&5.705&5.619&5.26&$5\cdot10^{-4}$&$6\cdot10^{-4}$&0.01&&\\
%&&13.439&13.515&11.295&&6.866&4.58&$1\cdot10^{-4}$&$1\cdot10^{-4}$&0.02&&\\
&&13.827&13.878&11.856&5.289&5.349&4.99&$1\cdot10^{-4}$&$1\cdot10^{-4}$&0.0016&&\\
&$^{20}$Ne($3^+$)&10.485&10.531&9.873&5.210&5.349&5.77&0.21&0.24&0.02&&\\
&&10.554&10.692&10.884&4.789&4.843&4.36&0.52&0.41&0.11&&\\
&$^{20}$Ne($1^+$)&11.166&12.754&11.262&3.871&3.850&3.72&2.07&1.88&0.2&&\\
$^{21}$Na(${3/2}^+$)&$^{21}$Ne($3/2^+$)&0&0&0&3.624&3.617&3.608&91.81&90.92&94.93&\cite{Wilson}&\\
&$^{21}$Ne($5/2^+$)&0.266&0.172&0.350&4.585&4.587&4.596&8.81&9.07&5.06&&\\
&$^{21}$Ne($1/2^+$)&2.859&2.679&2.794&4.764&4.754&4.61&$3\cdot10^{-4}$&$5\cdot10^{-4}$&--&&\\
$^{22}$Na(${3}^+$)&$^{22}$Ne($2^+$)&1.363&1.424&1.274&7.189&6.991&7.41&100&100&89.9&\cite{22na}&\\
$^{21}$Mg(${5/2}^+$)&$^{21}$Na($3/2^+$)&0&0&0&5.630&5.616&5.26&9.62&9.49&16&\cite{Sextro}&\\
&$^{21}$Na($5/2^+$)&0.266&0.172&0.331&4.787&4.777&4.79&60.19&61.21&41&&\\
&$^{21}$Na($7/2^+$)&1.757&1.691&1.716&4.806&4.812&5.11&30.18&29.28&10.9&&\\
$^{22}$Mg(${0}^+$)&$^{22}$Na($1^+$)&0.338&0.715&0.583&3.646&3.618&3.64&66.94&87.88&41.3&\cite{Achouri2}&\\
&&2.041&1.871&1.936&3.459&3.422&3.46&7.03&12.11&5.44&&\\
$^{23}$Mg(${3/2}^+$)&$^{23}$Na($3/2^+$)&0&0&0&3.694&3.692&3.667&88.14&86.73&92.14&\cite{Magron}&\\
&$^{23}$Na($5/2^+$)&0.399&0.332&0.440&4.423&4.413&4.434&11.08&13.18&7.84&&\\
&$^{23}$Na($1/2^+$)&2.173&2.062&2.390&4.717&4.714&4.97&0.028&0.08&0.006&&\\
$^{22}$Al(${(4)}^+$)&$^{22}$Mg($4^+$)&3.357&3.385&5.293&6.888&6.769&4.87&49.22&50.91&31&\cite{Achouri}&\\
&&5.367&5.276&6.221&4.807&4.775&5.33&4.28&4.18&7.4&&\\
%&&6.300&&6.307&5.734&5.712&5.51&9.71&&4.7&&\\
&$^{22}$Mg($3^+$)&5.461&5.506&5.451&5.283&5.290&5.56&15.94&14.48&5.9&&\\
&&6.513&6.535&6.865&6.376&6.264&5.6&0.91&0.91&3&&\\
&$^{22}$Mg($5^+$)&7.294&7.232&7.132&4.753&4.741&4.75&28.90&28.42&18.5&&\\
$^{23}$Al(${5/2}^+$)&$^{23}$Mg($3/2^+$)&0&0&0&5.671&5.693&5.3&29.42&23.82&36.3&\cite{23al}&\\
&$^{23}$Mg($5/2^+$)&0.399&0.332&0.451&5.630&5.559&5.36&27.21&28.08&26.2&&\\
&&3.750&3.685&7.803&7.490&7.167&3.31&0.07&0.11&13.4&&\\
&&5.422&5.319&8.164&5.667&5.675&4.77&1.28&1.17&0.28&&\\
&$^{23}$Mg($7/2^+$)&2.169&2.105&2.051&6.645&6.538&5.67&1.09&1.27&5.91&&\\
&&4.662&4.554&7.848&6.212&6.250&5.3&0.65&0.55&0.11&&\\
$^{24}$Al(${4}^+$)&$^{24}$Mg($4^+$)&4.372&4.336&4.122&5.899&5.805&6.13&100&100&7.7&\cite{Warburton}&\\
$^{25}$Al(${5/2}^+$)&$^{25}$Mg($5/2^+$)&0&0&0&3.588&3.582&3.57&99.25&99.22&99.25&\cite{Wilson}&\\
&$^{25}$Mg($3/2^+$)&1.097&1.122&0.974&6.092&6.028&6.2&0.06&0.06&0.07&&\\
&$^{25}$Mg($7/2^+$)&1.720&1.715&1.611&4.436&4.411&4.36&0.67&0.68&0.70&&\\
$^{25}$Si(${5/2}^+$)&$^{25}$Al($5/2^+$)&0&0&0&5.277&5.189&5.24&28.99&28.60&25&\cite{Thomas}&\\
&&1.995&1.993&3.844&6.148&6.036&6.21&1.60&1.67&0.4&&\\
&&3.908&4.033&4.582&6.791&6.769&5.11&0.12&0.196&3.2&&\\
&&4.740&4.749&5.802&5.090&5.057&5&3.79&3.27&1.7&&\\
&&5.799&5.817&6.620&5.805&4.290&5.7&0.33&0.005&0.16&&\\
&$^{25}$Al($3/2^+$)&1.097&1.122&0.945&5.069&5.020&5.05&29.22&26.05&26&&\\
&&2.811&2.879&2.672&5.102&5.000&5.42&11.68&11.48&4.8&&\\
&&4.332&4.369&4.189&5.437&5.440&5.25&1.59&1.74&2.99&&\\
&&5.788&5.781&6.170&5.330&5.317&5.6&1.01&0.85&0.32&&\\
&&6.346&6.334&7.107&5.465&5.440&4.16&0.40&0.47&3.7&&\\
&$^{25}$Al($7/2^+$)&1.720&1.715&1.613&5.156&5.151&5.16&17.89&14.64&15&&\\
$^{26}$Si(${0}^+$)&$^{26}$Al($1^+$)&0.806&1.088&1.057&3.562&3.569&3.54&55.71&52.91&21.9&\cite{Wilson}&\\
&&1.589&1.862&1.850&3.811&3.735&3.86&11.87&13.80&2.72&&\\
&&1.880&2.141&2.071&5.443&5.379&4.63&0.18&0.21&0.28&&\\
&&2.524&2.749&2.740&4.369&4.248&4.54&0.73&1.03&0.06&&\\
&&3.428&3.674&3.723&5.053&4.706&$>$4.2&0.018&0.043&$<$0.001&&\\
$^{27}$Si(${5/2}^+$)&$^{27}$Al($5/2^+$)&0&0&0&3.631&3.632&$>$3.6&99.52&99.33&$<$ 99.71&\cite{Daehnick}&\\
&&2.699&2.770&2.734&4.797&4.777&5&0.04&0.03&0.02&&\\
&$^{27}$Al($3/2^+$)&1.063&1.095&1.014&5.852&5.531&7.23&0.18&0.34&0.006&&\\
&&2.841&2.810&2.982&4.260&4.283&4.34&0.07&0.08&0.02&&\\
&$^{27}$Al($7/2^+$)&2.328&2.280&2.212&4.697&4.672&4.69&0.17&0.2&0.18&&\\
$^{26}$P(${(3)}^+$)&$^{26}$Si($2^+$)&1.897&1.982&1.797&4.773&4.771&4.9&60.67&59.66&44&\cite{Thomas}&\\
%%%%%%%%%%%%%%%%%%%%%%%%%%%%%%%%%%%%%%%%%%%%%%%%%%%%%%%%%%%%%%%%%%%%%%%%%%%%%%%%%%%%%%%%%%%%%%%%%%%%%%%%%%%%%%%%%%%%%%%%%%%%%%%                 

\hline 
\end{tabular}
\end{footnotesize}
\end{center}
\end{table*}
\pagestyle{plain}
%%%%%%%%%%%%%%%%%%%%%%%%%%%%%%%%%%%%%%%%%%%%%%%%%%%%%%%%%%%%%
%%%%%%%%%%%%%%%%%%%%%%%%%%%%%%%%%%%%%%%%%%%%%%%%%%%%%%%%%%%%%

\addtocounter{table}{-1}

\begin{table*}
  \begin{center}
\begin{footnotesize} 
    \leavevmode
    \caption{{\em Continuation.\/}}
\begin{tabular}{r@{\hspace{2pt}}c@{\hspace{0.1pt}}c@{\hspace{0.1pt}}c@{\hspace{0.1pt}}c@{\hspace{0.1pt}} c@{\hspace{0.1pt}}c@{\hspace{0.1pt}}c@{\hspace{0.1pt}}c@{\hspace{0.1pt}}c@{\hspace{0.1pt}}c@{\hspace{0.1pt}} c@{\hspace{0.5pt}}c@{\hspace{0.1pt}}}
\hline
&    &\multicolumn {3} {c} {Ex. energy (keV)} &\multicolumn {3} {c} {log$ft$ value}& \multicolumn {3} {c} {Branching ($\%$)}& Ref.\\
%\cline{3-5}
%\cline{6-8}
%\cline{9-11}
$^{A}$Z$_{i}(J^{\pi})$  & $^{A}$Z$_{f}(J^{\pi})$  &
   \multicolumn{1}{c}{USDB}& \multicolumn{1}{c}{SDNN}&\multicolumn{1}{c}{EXPT.} &
      \multicolumn{1}{c}{USDB}& \multicolumn{1}{c}{SDNN}&\multicolumn{1}{c}{EXPT.} &
      \multicolumn{1}{c}{USDB}& \multicolumn{1}{c}{SDNN}&\multicolumn{1}{c}{EXPT.}&
      \\
\hline
%%%%%%%%%%%%%%%%%%%%%%%%%%%%%%%%%%%%%%%%%%%%%%%%%%%%%%%%%%%%%%%%%%%%%%%%%%%%%%%%%%%%%%%%%%%%%%%%%%%%%%%%%%%%%%%%%%%%%%%%%%%%%%%%%%%

&&3.007&3.102&2.786&6.645&6.866&5.9&0.6&0.33&3.3&&\\
&&4.449&4.550&4.139&5.607&5.523&5.9&3.92&4.49&1.78&&\\
&&4.882&4.995&6.295&5.346&5.349&5.9&6.94&5.68&0.78&&\\
&&5.385&5.615&6.384&7.490&7.769&5.5&0.04&0.01&1.7&&\\
&&6.676&6.799&7.501&6.075&6.145&5.2&0.53&0.43&2.4&&\\
&$^{26}$Si($(3)^+$)&3.882&3.889&3.756&6.129&6.013&5.8&1.40&1.84&2.68&&\\
&&4.317&4.451&4.186&5.133&5.119&5.7&11.88&11.78&2.91&&\\
&&6.179&6.283&5.928&4.791&4.717&4.6&12.71&14.63&18&&\\
&$^{26}$Si($4^+$)&4.365&4.346&3.842&6.092&6.156&6&1.29&1.13&1.68&&\\
$^{28}$P(${3}^+$)&$^{28}$Si($2^+$)&1.932&2.004&1.779&4.886&4.899&4.851&73.18&70.01&69.1&\cite{28p}&\\
&$^{28}$Si($4^+$)&4.607&4.591&4.617&5.688&5.645&5.99&3.93&4.44&1.29&&\\
&$^{28}$Si($3^+$)&6.330&6.194&6.276&4.558&4.551&4.788&22.88&25.54&7.6&&\\
$^{29}$P(${1/2}^+$)&$^{29}$Si($1/2^+$)&0&0&0&3.706&3.702&3.681&96.74&96.62&98.29&\cite{Wilson}&\\
&$^{29}$Si($3/2^+$)&1.285&1.266&1.273&5.277&5.143&4.806&0.25&0.33&1.26&&\\
&&2.525&2.659&2.425&4.047&4.030&4.172&3.00&3.04&0.45&&\\
$^{30}$P(${1}^+$)&$^{30}$Si($0^+$)&0&0&0&4.919&4.950&4.839&99.92&99.81&99.93&\cite{Wilson}&\\
&&3.912&3.963&3.787&5.128&5.205&4.48&0.002&0.002&0.003&&\\
&$^{30}$Si($2^+$)&2.266&2.296&2.235&5.814&5.601&5.884&0.06&0.1&0.05&&\\
&&3.506&3.590&3.498&6.260&6.371&5.26&0.002&0.002&0.001&&\\
&$^{30}$Si($1^+$)&4.065&4.151&3.769&4.483&4.447&5.79&$1\cdot10^{-4}$&$1\cdot10^{-5}$&$1\cdot10^{-4}$&&\\
$^{29}$S(${(5/2)}^+$)&$^{29}$P($3/2^+$)&1.285&1.266&1.383&4.837&4.811&5.06&52.98&54.15&27&\cite{29s}&\\
&&2.525&2.659&2.422&5.230&5.182&4.98&13.10&13.21&20.7&&\\
&&6.047&6.091&9.389&6.376&6.468&4.4&0.16&0.123&0.43&&\\
&$^{29}$P($5/2^+$)&2.063&2.068&1.953&5.899&5.665&5.73&3.39&5.60&4.5&&\\
&&3.356&3.359&3.105&6.043&5.976&6.2&1.47&1.57&0.9&&\\
&&4.900&4.831&4.954&4.628&4.650&4.639&17.41&16.37&11.9&&\\
&$^{29}$P($7/2^+$)&4.222&4.188&4.080&7.791&7.070&$>$6.2&0.017&0.085&$<$ 0.5&&\\
&&5.122&5.156&5.293&4.952&4.841&5.03&7.33&8.88&3.9&&\\
$^{30}$S(${0}^+$)&$^{30}$P($1^+$)&0&0&0&4.443&4.473&4.322&25.26&22.88&21.3&\cite{Wilson}&\\
&&0.647&0.660&0.708&4.862&4.756&5.89&5.14&6.31&0.29&&\\
&&3.138&3.285&3.019&3.608&3.531&3.549&2.44&2.03&2.28&&\\
$^{31}$S(${1/2}^+$)&$^{31}$P($1/2^+$)&0&0&0&3.687&3.688&3.678&98.36&98.06&98.85&\cite{Wilson}&\\
&&3.236&3.261&3.134&4.917&4.851&4.76&0.02&0.02&0.03&&\\
&$^{31}$P($3/2^+$)&1.173&1.133&1.266&5.005&4.939&4.96&1.60&1.90&1.1&&\\
&&3.625&3.772&3.506&4.371&4.359&4.57&0.01&0.007&0.01&&\\
$^{31}$Cl(${3/2}^+$)&$^{31}$S($1/2^+$)&0&0&0&5.517&5.481&5.6&10.74&11.02&7&\cite{Bennett}&\\
&&3.237&3.261&3.076&5.471&5.546&5.33&2.25&1.76&2.54&&\\
&$^{31}$S($3/2^+$)&1.174&1.133&1.248&5.285&5.235&6.1&10.64&11.54&1.1&&\\
&&3.626&3.772&3.434&6.449&6.145&5.84&0.18&0.31&0.64&&\\
&&4.375&4.457&4.207&4.843&5.756&4.81&4.5&3.52&4.1&&\\
&$^{31}$S($5/2^+$)&2.306&2.275&2.234&4.280&4.263&4.37&59.97&60.04&38&&\\
&&3.313&3.280&3.283&5.018&4.999&5.03&6.09&6.14&4.46&&\\
$^{33}$Cl(${3/2}^+$)&$^{33}$S($3/2^+$)&0&0&0&3.732&3.730&3.747&95.48&98.30&98.58&\cite{Wilson}&\\
&&2.241&2.220&2.312&5.823&5.756&5.82&0.05&0.06&0.04&&\\
&&3.585&3.615&3.935&5.972&6.013&$>$6.3&0.001&$8\cdot10^{-4}$&$<$$1\cdot10^{-4}$&&\\
&$^{33}$S($1/2^+$)&0.852&0.831&0.840&5.674&5.616&5.65&0.63&0.7&0.48&&\\
&&3.943&3.996&4.053&5.404&5.376&6.2&$7\cdot10^{-4}$&$6\cdot10^{-4}$&$1\cdot10^{-4}$&&\\
&&4.277&4.299&4.375&4.987&4.955&5.7&$4\cdot10^{-4}$&$4\cdot10^{-4}$&$<$$1\cdot10^{-4}$&&\\
&$^{33}$S($5/2^+$)&1.923&1.889&1.966&4.264&5.183&4.97&0.35&0.42&0.46&&\\
&&2.949&2.906&2.867&4.291&4.228&4.18&0.39&0.48&0.43&&\\
&&3.842&3.856&3.832&4.959&4.958&$>$6.4&0.003&0.003&$<$0.0001&&\\
$^{34}$Cl(${3}^+$)&$^{34}$S($2^+$)&2.131&2.135&2.127&5.867&6.028&5.982&16.78&26.34&28.52&\cite{34cl}&\\
&&3.123&3.133&3.304&4.719&4.680&4.823&23.12&23.33&26.42&&\\
&&4.126&4.186&4.114&4.933&4.880&5.081&0.37&0.38&0.45&&\\
&$^{34}$S($4^+$)&4.836&4.790&4.688&5.723&5.642&5.44&39.99&46.718&--&&\\
&$^{34}$S($3^+$)&4.702&4.682&4.876&5.264&5.187&5.18&10.17&12.78&--&&\\
$^{32}$Ar(${0}^+$)&$^{32}$Cl($1^+$)&0&0&0&5.959&5.905&$>$5.1&1.13&1.19&$<$7&\cite{Bhattacharya}&\\
&&1.119&1.082&1.168&3.874&3.854&3.94&78.51&78.39&57&&\\
&&1.981&1.958&2.210&5.125&5.110&5.9&2.73&2.67&0.35&&\\
&&2.875&2.861&3.772&4.614&4.618&4.424&5.11&4.77&3.68&&\\
&&3.806&3.825&4.167&5.262&5.289&4.9&0.6&0.52&0.9&&\\
$^{33}$Ar(${1/2}^+$)&$^{33}$Cl($3/2^+$)&0&0&0&5.071&5.020&5.022&21.19&22.76&18.7&\cite{33ar}&\\
&&2.241&2.220&2.352&4.977&4.980&5.54&8.15&8.50&1.7&&\\
&&3.585&3.615&3.971&5.651&5.546&5.744&0.94&0.79&0.38&&\\

%%%%%%%%%%%%%%%%%%%%%%%%%%%%%%%%%%%%%%%%%%%%%%%%%%%%%%%%%%%%%%%%%%%%%%%%%%%%%%%%%%%%%%%%%%%%%%%%%%%%%%%%%%%%%%%%%%%%%%%%%%%%%%%%%%%

 \hline
\end{tabular}
\end{footnotesize}
\end{center}
\end{table*}
%%%%%%%%%%%%%%%%%%%%%%%%%%%%%%%%%%%%%%%%%%%%%%%%%%%%%%%%%%%%%%%%%%%%%%
%%%%%%%%%%%%%%%%%%%%%%%%%%%%%%%%%%%%%%%%%%%%%%%%%%%%%%%%%%%%%%%%%%%%%%
\addtocounter{table}{-1}

\begin{table*}
  \begin{center}
\begin{footnotesize} 
    \leavevmode
    \caption{{\em Continuation.\/}}
\begin{tabular}{r@{\hspace{2pt}}c@{\hspace{0.1pt}}c@{\hspace{0.1pt}}c@{\hspace{0.1pt}}c@{\hspace{0.1pt}} c@{\hspace{0.1pt}}c@{\hspace{0.1pt}}c@{\hspace{0.1pt}}c@{\hspace{0.1pt}}c@{\hspace{0.1pt}}c@{\hspace{0.1pt}} c@{\hspace{0.5pt}}c@{\hspace{0.1pt}}}
\hline
&    &\multicolumn {3} {c} {Ex. energy (keV)} &\multicolumn {3} {c} {log$ft$ value}& \multicolumn {3} {c} {Branching ($\%$)}& Ref.\\
%\cline{3-5}
%\cline{6-8}
%\cline{9-11}
$^{A}$Z$_{i}(J^{\pi})$  & $^{A}$Z$_{f}(J^{\pi})$  &
   \multicolumn{1}{c}{USDB}& \multicolumn{1}{c}{SDNN}&\multicolumn{1}{c}{EXPT.} &
      \multicolumn{1}{c}{USDB}& \multicolumn{1}{c}{SDNN}&\multicolumn{1}{c}{EXPT.} &
      \multicolumn{1}{c}{USDB}& \multicolumn{1}{c}{SDNN}&\multicolumn{1}{c}{EXPT.}&
      \\
\hline
%%%%%%%%%%%%%%%%%%%%%%%%%%%%%%%%%%%%%%%%%%%%%%%%%%%%%%%%%%%%%%%%%%%%%%%%%%%%%%%%%%%%%%%%%%%%%%%%%%%%%%%%%%%%%%%%%%%%%%%%%%%%%%%%%%%

&&4.355&4.379&4.113&4.916&4.915&5.626&2.47&2.32&0.45&&\\
&$^{33}$Cl($1/2^+$)&0.852&0.831&0.810&4.459&4.452&4.516&56.99&58.09&41&&\\
&&3.943&3.996&4.441&6.615&7.292&4.8&0.06&0.01&2.37&&\\
&&4.277&4.299&5.548&4.570&4.538&3.284&5.8&5.87&31&&\\
&&5.217&5.220&5.732&4.521&4.514&5.92&3.07&2.92&0.06&&\\
$^{34}$Ar(${0}^+$)&$^{34}$Cl($1^+$)&0.325&0.361&0.460&5.940&5.991&5.31&0.63&0.53&0.91&\cite{Iacob}&\\
&&0.512&0.543&0.665&4.413&4.405&4.78&17.64&17.24&2.49&&\\
&&2.372&2.420&2.579&4.596&4.403&4.12&1.30&1.88&0.86&&\\
&&3.031&3.069&3.129&3.615&3.578&3.458&4.23&4.12&1.29&&\\
$^{35}$Ar(${3/2}^+$)&$^{35}$Cl($3/2^+$)&0&0&0&3.742&3.740&3.752&97.29&97.24&98.23&\cite{Adelberger}&\\
&&2.625&2.638&2.645&5.015&4.963&$>$5.8&0.24&0.26&$<$0.03&&\\
&&3.946&4.024&2.693&4.949&4.911&4.989&0.008&0.006&0.16&&\\
&&4.765&4.833&3.918&5.235&5.174&4.88&0.001&$1\cdot10^{-4}$&0.01&&\\
&$^{35}$Cl($1/2^+$)&1.225&1.237&1.219&5.127&5.088&5.091&1.59&1.63&1.23&&\\
&&3.980&4.055&3.967&4.839&4.830&4.81&0.009&0.006&0.007&&\\
&$^{35}$Cl($5/2^+$)&1.677&1.657&1.763&5.167&5.160&5.475&0.81&0.8&0.24&&\\
&&3.096&3.092&3.002&5.395&5.344&4.973&0.03&0.04&0.08&&\\
$^{35}$K(${(3/2)}^+$)&$^{35}$Ar($3/2^+$)&0&0&0&4.942&4.920&5.07&37.49&37.59&19&\cite{Ewan}&\\
&&2.625&2.638&2.638&6.360&7.167&$\ge$6.2&0.34&0.41&$\le$0.4&&\\
&&3.946&4.024&4.065&6.360&7.167&5.64&0.03&0.02&0.55&&\\
&$^{35}$Ar($1/2^+$)&1.225&1.237&1.184&7.791&7.769&5.77&0.02&0.02&2.2&&\\
&&3.980&4.055&4.528&4.876&4.876&4.85&4.56&4.18&0.7&&\\
&$^{35}$Ar($5/2^+$)&1.677&1.657&1.750&4.787&4.782&4.91&22&21.71&11.9&&\\
&&3.096&3.092&2.982&4.302&4.266&4.27&30.3&31.86&26&&\\
$^{36}$K(${2}^+$)&$^{36}$Ar($2^+$)&1.818&1.814&1.970&4.671&4.659&4.78&73.02&74.32&44&\cite{ENSDF}&\\
&&4.254&4.319&4.441&4.662&4.634&4.9&25.10&25.13&8.4&&\\
&&6.375&6.441&4.950&5.862&5.769&6.48&0.47&0.54&0.16&&\\
$^{37}$K(${3/2}^+$)&$^{37}$Ar($3/2^+$)&0&0&0&3.655&3.646&3.657&97.21&96.87&97.88&\cite{Hagberg}&\\
&&3.653&3.693&3.601&4.913&4.889&4.958&0.02&0.02&0.02&&\\
&&4.845&4.857&3.937&5.357&5.347&5.8&$1\cdot10^{-4}$&$1\cdot10^{-4}$&0.001&&\\
&$^{37}$Ar($1/2^+$)&1.386&1.362&1.409&6.712&6.655&7.38&0.03&0.02&0.004&&\\
&$^{37}$Ar($5/2^+$)&2.771&2.756&2.796&3.900&3.797&3.785&2.22&2.71&2.06&&\\
&&3.155&3.129&3.169&4.201&4.356&6.34&0.51&0.35&0.002&&\\
$^{38}$K(${3}^+$)&$^{38}$Ar($2^+$)&1.843&1.850&2.167&4.707&4.684&4.975&92.65&92.64&99.84&\cite{38k}&\\
&&4.239&4.221&3.935&6.158&6.237&5.88&0.05&0.04&0.14&&\\
&&9.935&10.070&4.565&5.566&5.551&$>$4.7&0.011&0.011&$<$0.15&&\\
%&&12.086&12.015&5.157&&7.769&$>$5.0&$1\cdot10^{-5}$&$1\cdot10^{-5}$&$<$0.02&&\\
%&$^{38}$Ar($3^+$)&10.578&10.324&3.810&&&$>$7.3&$1\cdot10^{-5}$&$1\cdot10^{-5}$&$<$0.01&&\\
&$^{38}$Ar($4^+$)&8.759&8.668&5.349&3.484&3.461&$>$4.6&7.29&7.29&$<$0.02&&\\
$^{36}$Ca(${0}^+$)&$^{36}$K($1^+$)&1.071&1.063&1.112&4.040&4.039&4.519&47.24&44.59&14.3&\cite{36ca}&\\
&&1.732&1.732&1.618&4.044&3.975&4.06&33.73&37.06&31&&\\
&&2.553&2.578&3.357&3.845&4.879&4.06&3.43&2.92&10.3&&\\
&&3.379&3.376&4.450&4.071&4.058&4.29&12.54&12.11&2.6&&\\
&&5.591&5.634&4.658&4.672&4.580&4.55&0.63&0.7&1.2&&\\
&&6.124&6.111&5.926&3.952&3.888&3.73&2.05&2.25&2.2&&\\
&&6.547&6.527&6.787&4.554&4.563&3.900&0.34&0.33&0.5&&\\
$^{37}$Ca(${3/2}^+$)&$^{37}$K($3/2^+$)&0&0&0&4.892&4.890&5.040&27.68&26.83&18.5&\cite{Kaloskamis}&\\
&&3.653&3.693&3.622&4.708&4.648&4.930&5.74&6.18&3.3&&\\
&$^{37}$K($1/2^+$)&1.386&1.362&1.370&5.092&5.082&5.690&8.94&8.97&2.14&&\\
&&4.265&4.287&4.192&4.063&4.043&6.330&16.47&16.39&0.09&&\\

%&&5.333&5.343&4.496&&&5.050&0.86&0.8&1.34&&\\
&$^{37}$K($5/2^+$)&2.771&2.756&2.750&4.161&4.170&4.790&35.38&33.82&7.9&&\\
&&3.155&3.129&3.239&5.799&5.254&4.900&0.64&2.21&4.5&&\\
$^{38}$Ca(${0}^+$)&$^{38}$K($1^+$)&0.538&0.546&0.458&4.919&4.906&4.800&3.686&3.51&2.84&\cite{Blank}&\\
&&1.504&1.464&1.697&3.450&3.417&3.426&46.02&48.22&19.47&&\\
&&4.221&4.201&3.341&3.859&3.865&4.160&0.40&0.39&0.34&&\\
&&5.578&5.573&3.855&7.803&7.789&4.800&$1\cdot10^{-5}$&$1\cdot10^{-5}$&0.02&&\\
&&6.665&6.587&3.976&3.778&3.780&4.050&$6\cdot10^{-4}$&$1\cdot10^{-5}$&0.11&&\\
&&11.026&10.906&4.174&3.576&3.566&5.300&$1\cdot10^{-5}$&$1\cdot10^{-5}$&$1\cdot10^{-4}$&&\\
$^{39}$Ca(${3/2}^+$)&$^{39}$K($3/2^+$)&0&0&0&3.596&3.591&3.630&99.96&99.96&99.92&\cite{39ca}&\\
&$^{39}$K($1/2^+$)&2.648&2.633&2.522&3.799&3.799&7.0004&0.0004&0&0.002&&\\

%%%%%%%%%%%%%%%%%%%%%%%%%%%%%%%%%%%%%%%%%%%%%%%%%%%%%%%%%%%%%%%%%%%%%%%%%%%%%%%%%%%%%%%%%%%%%%%%%%%%%%%%%%%%
 \hline
\end{tabular}
\end{footnotesize}
\end{center}
\end{table*}
%%%%%%%%%%%%%%%%%%%%%%%%%%%%%%%%%%%%%%%%%%%%%%%%%%%%%%%%%%%%%%%%%%%%%%%%%%%%%%%%%%%%%%%%%%%%%%%%%%%%%%
\begin{table*}
\begin{center}
    \leavevmode
    \caption{\label{table4}The superallowed transitions($0^+$ $\rightarrow$ $0^+$) are listed for the concerned nuclei. The theoretical log$ft$ values and branching ratios of $\beta^+$/EC decay of the concerned nuclei are compared with the experimental data. The experimental ground state energy and parity $J^\pi$ of the parent and daughter nucleus are listed along with $Q$ values; the references to the experimental data are given in the last column; 
 $J_{P}^{\pi}$ ( $J_{D}^{\pi}$) and $E_P$ ($E_D$) are spin-parity and excitation energy of parent (daughter) nuclei, respectively.  }
\begin{tabular}{lrccccccccccccc}
\hline
  & & & & & & & \multicolumn{3}{c}{log$ft$}& &\multicolumn{3}{c}{Branch (\%)} & \\
\cline{8-10}
\cline{12-14}
Nuclide & Decay  & Q(keV) & $E_P$(keV) & $J_{P}^{\pi}$ & $E_D$(keV) & $J_{D}^{\pi}$ & \multicolumn{1}{c}{USDB}&\multicolumn{1}{c}{SDNN}&\multicolumn{1}{c}{Expt.}&& \multicolumn{1}{c}{USDB}&\multicolumn{1}{c}{SDNN}&\multicolumn{1}{c}{Expt.} & Ref.\\
%\cline{5-6}
\hline
$^{18}$Ne   & $\beta^+$/EC &  4445.7  & 0   & $0^+$ &  1041.5 &  $0^+$ &3.497 &3.497 & 3.468 && 4.95 & 4.95  &7.69 & \cite{18ne}\\
$^{22}$Mg   & $\beta^+$/EC &  4781.6  & 0   & $0^+$ &  657    &  $0^+$ & 3.497&3.497 & 3.487 && 26.2 &  25.25 & 53.21& \cite{Achouri2}\\
$^{26}$Si   & $\beta^+$/EC & 5069.14  & 0   & $0^+$ &  228.3  &  $0^+$ & 3.497&3.497 & 3.490 && 31.48 &  31.99 &75.05 & \cite{Hardy}\\
$^{30}$S    & $\beta^+$/EC & 6138     & 0   & $0^+$ &  677.01 &  $0^+$ & 3.523&3.523 & 3.485 && 67.15 & 68.77 &76.14 & \cite{Wilson}\\
$^{32}$Ar   & $\beta^+$/EC & 11134.7  & 0   & $0^+$ &  5046.3  &  $0^+$ & 7.507&7.507 & 3.179&& 0.02 & 0.014  & 22.70& \cite{Bhattacharya}\\
$^{34}$Ar   & $\beta^+$/EC & 6061.83  & 0   & $0^+$ &  0.0    &  $0^+$ & 3.636&3.636 & 3.485 && 76.17 &  76.11 &94.46 & \cite{Iacob}\\
%$^{36}$Ca   & $\beta^+$/EC & 10966    & 0.0   & $0^+$ &  4281.9 &  $0^+$ & 13.762&13.762 & 3.184&& 2.73 &  -&38.02 & \cite{36ca}\\
$^{38}$Ca   & $\beta^+$/EC & 6742.26  & 0   & $0^+$ &  130.2  &  $0^+$ & 3.497&3.497 & 3.487 && 49.87 &  47.85&77.19 & \cite{Blank}\\
\hline
\end{tabular}
\end{center}
\end{table*}
%%%%%%%%%%%%%%%%%%%%%%%%%%%%%%%%%%%%%%%%%%%%%%%%%%%%%%%%%%%%%%%%%%%%%%%%%%%%%%%%%%%%
\begin{table*}
  \begin{center}
  \begin{small}  
    \leavevmode
    \caption{\label{table5}The theoretical $\beta^+$/EC -decay half-lives are compared
    with the experimental data for the concerned nuclei; the experimental
$Q$ values; {$\beta^+ + \epsilon$} -decay probabilities; quenched theoretical sum $B(GT)$ values.}

\begin{tabular}{lrccccccccccc}
\hline
   &    & $Q$ value & \multicolumn{2}{c}{Sum $B(GT)$}& \multicolumn{3}{c}{$\beta$-decay half-life}& ${\beta^+ + \epsilon}$\\
%\cline{4-5}
%\cline{6-7}
$^{A}$Z$_{i}(J^{\pi})$  & $^{A}$Z$_{f}$  & (keV) & \multicolumn{1}{c}{USDB}&\multicolumn{1}{c}{SDNN}&\multicolumn{1}{c}{USDB}&\multicolumn{1}{c}{SDNN}&\multicolumn{1}{c}{EXPT.}& $\%$\\
\hline
$^{18}$Ne(${0}^+$)&$^{18}$F&4445.7$\pm$47&3.229&3.336&1.645 s&1.559 s&1.672$\pm$8 s\cite{18ne}&100&\\
$^{19}$Ne(${1/2}^+$)&$^{19}$F&3238.4$\pm$6&1.693&1.753&18.224 s&17.252 s&17.22$\pm$2 s\cite{19ne}&100&\\
$^{20}$Na(${2}^+$)&$^{20}$Ne&13886$\pm$7&1.583&1.606&221.6 ms&215.02 ms&447.9$\pm$23 ms\cite{ENSDF}&100&\\
$^{21}$Na(${3/2}^+$)&$^{21}$Ne&3547.14$\pm$28&0.462&0.469&29.568 s&27.756 s&22.49$\pm$4 s\cite{Wilson}&100&\\
$^{22}$Na(${3}^+$)&$^{22}$Ne&2843.20$\pm$17&0.00024&0.00038&1.57 y&0.99 y&2.6018$\pm$22 y\cite{22na}&100&\\
$^{21}$Mg(${5/2}^+$)&$^{21}$Na&13098$\pm$16&0.130&0.128&170.8 ms&163.02 ms&122$\pm$3 ms\cite{Sextro}&100&\\
$^{22}$Mg(${0}^+$)&$^{22}$Na&4781.6$\pm$3&2.174&2.305&2.93 s&2.696 s&3.875$\pm$12 s\cite{Achouri2}&100&\\
$^{23}$Mg(${3/2}^+$)&$^{23}$Na&4056.17$\pm$32&0.380&0.381&16.38 s&15.69 s&11.304$\pm$45 s\cite{Magron}&100&\\
$^{22}$Al(${(4)}^+$)&$^{22}$Mg&17560&0.155&0.159&56.34 ms&50.71 ms&91.1$\pm$5 ms\cite{Achouri}&100&\\
$^{23}$Al(${5/2}^+$)&$^{23}$Mg&12221.6$\pm$4&1.449&1.661&813.69 ms&715.6 ms&446$\pm$6 ms\cite{23al}&100&\\
$^{24}$Al(${4}^+$)&$^{24}$Mg&13884.77$\pm$4&0.005&0.006&2.481 s&1.994 s&2.053$\pm$4 s\cite{Warburton}&100&\\
$^{25}$Al(${5/2}^+$)&$^{25}$Mg&4276.7$\pm$5&0.519&0.531&9.63 s&9.206 s&7.183$\pm$12 s\cite{Wilson}&100&\\
$^{25}$Si(${5/2}^+$)&$^{25}$Al&12740$\pm$10&1.705&0.879&238.7 ms&218.7 ms&220$\pm$3 ms\cite{Thomas}&100&\\
$^{26}$Si(${0}^+$)&$^{26}$Al&5069.14$\pm$8&1.834&1.987&1.495 s&1.441 s&2.245$\pm$7 s\cite{Wilson}&100&\\
$^{27}$Si(${5/2}^+$)&$^{27}$Al&4812.36$\pm$10&0.635&0.624&5.63 s&5.46 s&4.16$\pm$4 s\cite{Daehnick}&100&\\
$^{26}$P(${(3)}^+$)&$^{26}$Si&18258$\pm$90&0.190&0.200&43.13 s&42.11 s&43.7$\pm$6 ms\cite{28p}&100&\\
$^{28}$P(${3}^+$)&$^{28}$Si&14345.1$\pm$12&0.162&0.160&154.45 ms&152.02 ms&270.3$\pm$5 ms\cite{28p}&100&\\
$^{29}$P(${1/2}^+$)&$^{29}$Si&4942.5$\pm$6&0.503&0.521&9.37 s&9.256 s&8.862$\pm$15 s\cite{Wilson}&100&\\
$^{30}$P(${1}^+$)&$^{30}$Si&4232.4$\pm$3&0.207&0.209&3.04  min&3.26 min&2.498$\pm$4 min\cite{Wilson}&100&\\
$^{29}$S(${5/2}^+$)&$^{29}$P&13795$\pm$50&0.219&0.232&119.69 ms&126.79 ms&188$\pm$4 ms\cite{29s}&100&\\
$^{30}$S(${0}^+$)&$^{30}$P&6138$\pm$3&1.123&1.286&1.813 s&1.761 s&1.178$\pm$5 s\cite{Wilson}&100&\\
$^{31}$S(${1/2}^+$)&$^{31}$P&5398.01$\pm$23&0.422&0.430&3.575 s&3.441 s&2.553$\pm$18 s\cite{Wilson}&100&\\
$^{31}$Cl(${3/2}^+$)&$^{31}$S&12008$\pm$3&0.442&0.402&257.19 ms&251.94 ms&190$\pm$1 ms\cite{Bennett}&100&\\
$^{33}$Cl(${3/2}^+$)&$^{33}$S&5582.59$\pm$44&0.614&0.462&3.387 s&3.322 s&2.511$\pm$3 s\cite{Wilson}&100&\\
$^{34}$Cl(${3}^+$)&$^{34}$S&5491.63$\pm$43&0.150&0.163&29.89 min&27.35 min&31.99$\pm$3 min\cite{34cl}&100&\\
$^{32}$Ar(${0}^+$)&$^{32}$Cl&11134.7$\pm$20&3.695&4.066&111.81 ms&104.24 ms&98$\pm$2 ms\cite{Bhattacharya}&100&\\
$^{33}$Ar(${1/2}^+$)&$^{33}$Cl&11619.1$\pm$6&0.476&0.483&217.67 ms&206.87 ms&173.0$\pm$20 ms\cite{33ar}&100&\\
$^{34}$Ar(${0}^+$)&$^{34}$Cl&6061.83$\pm$8&1.168&1.280&1580.68 ms&1497.39 ms&843.8$\pm$4 ms\cite{Iacob}&100&\\
$^{35}$Ar(${3/2}^+$)&$^{35}$Cl&5966.1$\pm$7&0.310&0.324&2.501 s&2.383 s&1.775$\pm$7 s\cite{Adelberger}&100&\\
$^{35}$K(${(3/2)}^+$)&$^{35}$Ar&11874.5$\pm$9&0.348&0.358&258.92 ms&246.47 ms&178$\pm$8 ms\cite{Ewan}&100&\\
$^{36}$K(${2}^+$)&$^{36}$Ar&12814.21$\pm$35&0.205&0.188&181.66 ms&177.91 ms&342$\pm$2 ms\cite{ENSDF}&100&\\
$^{37}$K(${3/2}^+$)&$^{37}$Ar&6147.48$\pm$23&1.017&1.073&1.624 s&1.523 s&1.225$\pm$7 s\cite{Hagberg}&100&\\
$^{38}$K(${3}^+$)&$^{38}$Ar&5914.07$\pm$4&1.333&1.376&3.87  min&3.67 min&7.651$\pm$19 min\cite{38k}&100&\\
$^{36}$Ca(${0}^+$)&$^{36}$K&10966$\pm$40&2.165&1.795&46.94 ms&59.37 ms&101.2$\pm$20 ms\cite{36ca}&100&\\
$^{37}$Ca(${3/2}^+$)&$^{37}$K&11664.5$\pm$8&0.953&0.995&184.99 ms&176.06 ms&181.1$\pm$10 ms\cite{Kaloskamis}&100&\\
$^{38}$Ca(${0}^+$)&$^{38}$K&6742.26$\pm$6&3.655&3.779&493.30 ms&448.71 ms&443.76$\pm$35 ms\cite{Blank}&100&\\
$^{39}$Ca(${3/2}^+$)&$^{39}$K&6524.5$\pm$6&1.839&1.901&1007.24 ms&954.49ms&860.3$\pm$10 ms\cite{39ca}&100&\\

\hline
\end{tabular}
\end{small}  
\end{center}
\end{table*}
%%%%%%%%%%%%%%%%%%%%%%%%%%%%%%%%%%%%%%%%%%%%%%%%%%%%%%%%%%%%%%%%%%%%%%%%%%%%
\section{Results and discussions}
We have compared the experimental matrix elements $M(GT)$ of 116 pure Gamow-Teller decays with the theoretical shell-model results using USDB and SDNN interactions in Table \ref{table1}. In most cases, the computed values are clearly systematically larger than the experimental ones. 
We have plotted the experimental values versus the theoretical ones for $R(GT)$ in Fig.~\ref{fig1}. Each Gamow-Teller transition is represented by a point in Fig.~\ref{fig1}. 
The points follow a straight line whose slope gives the average quenching factor, $q = 0.794\pm0.05 $ for USDB interaction and $q = 0.815\pm0.04 $ for SDNN interaction. 
 The eight individual Gamow-Teller transition strengths of $^{33}$Ar have been used to calculate the quenching factor 0.700$\pm$0.002 for $sd$ space using USD interaction in \cite{33ar}. In \cite{Thomas}, the quenching factor 0.6 is reported for $sd$ space using USD interaction.
In the previous work by Wildenthal {\it et al.} \cite{wildenthal3}, where $q = 0.77\pm0.02 $ has been reported for selected $sd$ shell nuclei. The calculated quenching factor in the present work is slightly greater than the previous works \cite{33ar}, \cite{Thomas}, \cite{wildenthal3}. Our calculated average quenching factors are more closer to unity.

Table \ref{table2} contains the calculated phase space factors for $\beta^+$ and EC decay only for those nuclei that have significant EC phase space factors. The table reflects that the EC phase space factors play an essential role in the decay process where the $Q$-values are equal to or less than 2 MeV. In this work, we have included the $\beta^+$/EC decay phase space factor in the half-life calculations. 

The total $\beta^+$/EC -decay half-life is calculated using the partial half-lives of all the transitions. So, to better understand the nuclear $\beta$ -decay, we have to analyze the details of each transition.
The experimental data with theoretical shell-model results of excitation energies and $\beta^+$-decay properties like log$ft$ values, branching percentages of 37 proton-rich and $N = Z$ nuclei of $sd$ shell are compared in Table \ref{table3}. The initial and final nuclei and the spin parity, are in columns 1 and 2, respectively.
The theoretical results using USDB, SDNN and experimental excitation energies (in keV) of each state involved in $\beta^+$-decays are reported in columns 3, 4, and 5, respectively. Columns 6, 7 and 8 reported the theoretical results using USDB, SDNN and experimental log$ft$ values, respectively. The Eq.~\ref{eq:15} has been used to calculate theoretical branching fractions. Columns 9 and 10 represent theoretical results using USDB and SDNN interactions, respectively, and the experimental values are listed in column 11. The rms deviation of the excitation energy between theory and experiment is 1281 keV, and the average relative error is 13.06$\%$ for USDB interaction, and for SDNN interaction, the rms deviation is 1362 keV with an average relative error of 14.06$\%$. Overall, the excitation energies predicted by USDB interaction are closer to the experimental data than those predicted by SDNN interaction. 

 In \cite{wildenthal4}, the log$ft$ values are reported using shell-model for the $\beta^+$-decay of $^{37}$K nucleus, the log$ft$ values are 3.53, 6.32, and 3.41 for the transitions at $3/2^+$, $1/2^+$, and $5/2^+$, respectively. In the present work, the SM results are 3.655, 6.712, and 3.9 using USDB and 3.646, 6.655, and 3.797 using SDNN interactions for the transitions at $3/2^+$, $1/2^+$, and $5/2^+$, respectively. Our theoretical quenched log$ft$ values are closer to the experimental values 3.657, 7.38, and 3.785 for the transitions at $3/2^+$, $1/2^+$, and $5/2^+$, respectively.
The observed weak branching fractions in $^{20}$Na, $^{30}$P, $^{33}$Cl, $^{37}$K, $^{38}$K, and $^{38}$Ca are successfully reproduced by shell model calculation using USDB and SDNN interactions. Overall, the shell model results for $sd$ shell nuclei agree with the experimental data for excitation energies, log$ft$ values and the branching ratios. There are a few cases where the theoretical results deviated from the experimental data.

The seven superallowed transitions with $0^+$ $\rightarrow$ $0^+$ for $^{18}$Ne, $^{22}$Mg, $^{26}$Si, $^{30}$S, $^{32}$Ar, $^{34}$Ar, and $^{38}$Ca are reported in Table \ref{table4}. We have also included the superallowed transitions in the $\beta$-decay half-lives calculation. From Table \ref{table4}, the theoretical results of log$ft$ values are in good agreement with the experimental data except for $^{32}$Ar. In the case of $^{32}$Ar, the calculated  log$ft$ values are larger than the experimental one; this is because the calculated Fermi matrix element is small; hence, Fermi reduced transition probability $B(F)$ is small, and hence log$ft$ value is large. Due to the large log$ft$ value, the branching ratio is smaller in this case. Overall, the branching ratios for the superallowed transitions agree with the experimental data.

 The theoretical and the experimental $\beta$-decay half-lives of the concerned nuclei are compared in Table \ref{table5}. The $\beta^+$/EC -decay half-lives are displayed in the log frame in Fig.\ref{fig2}. The experimental $Q$ -values, $\beta^+$+$\epsilon$ (electron capture) decay probabilities and the quenched theoretical sum $B(GT)$ values are also reported in Table \ref{table5}. For all 37 nuclei, their observed probabilities of $\beta^+$+$\epsilon$ -decay are 100$\%$.
The parent and daughter nuclei are listed in the first and second columns. Column 3 presents the experimental $Q$ -values taken from \cite{ENSDF}. The sums of quenched $B(GT)$ values are presented in columns 4 and 5. The theoretical (quenched) $\beta$-decay half-lives are presented in columns 6 and 7 for USDB and SDNN interactions, respectively, while the experimental data are given in column 8. 
The last column represents the experimental probabilities of $\beta^+$+$\epsilon$ decay.
The average relative error of the theoretical $\beta^+$/EC -decays half-lives is 33.1$\%$ for USDB interaction while 29.3$\%$ for SDNN interaction. For Ne, both the interactions predict the relative error less than 5$\%$, while other nuclei exhibit higher relative errors. The relative errors for nuclei such as $^{18}$Ne, $^{19}$Ne, $^{24}$Al, $^{25}$Si, $^{26}$P, $^{29}$P, $^{34}$Cl, $^{32}$Ar, $^{37}$Ca, $^{38}$Ca, and $^{39}$Ca are all below or equal to 20$\%$. For $^{34}$Ar, $^{36}$K, and $^{36}$Ca, the theoretical results are quite different from the experimental data. It might be because of two reasons: either the ground state angular momentum is not predicted correctly, or the used $Q$ -values have large uncertainties. This leads to errors in the theoretical half-lives calculations. A comparative analysis of $\beta^+$/EC -decay half-lives between USDB and SDNN interactions show that the theoretical results for the SDNN interaction with a quenching factor of 0.815 exhibits closer results than those of the USDB interaction, which has a quenching factor of 0.794.

%%%%%%%%%%%%%%%%%%%%%%%%%%%%%%%%%%%%%%%%%%%%%%

 We included the electron capture phase space factor using Eq.~\ref{eq:12}, and ~\ref{eq:13} in the calculations of beta decay half-lives.
For instance, in the case of $^{29}$P, the initial half-lives were calculated to be 9.265 and 8.871 s. After incorporating EC, these values changed to 9.256 and 8.863 s, respectively. This indicates that the results from SDNN interaction is closer to the experimental half-life 8.862 s than USDB interaction.
The influence of the EC phase space factor is relatively small in the cases of $^{22}$Na, $^{23}$Mg, $^{30}$P, $^{34}$Cl, and  $^{38}$K.
In the absence of electron capture, the half-life of $^{22}$Na are 1.75 and 1.10 y for USDB and SDNN interactions, respectively. However, after considering the EC phase space factor, the half-life become 1.58 and 0.98 y for USDB and SDNN interactions, respectively, while the experimental data is 2.6018 y. 
Excluding the EC phase space factor, the half-life of $^{23}$Mg was determined to be 16.39 and 15.70 s for the USDB and SDNN, respectively when the EC factor was taken into account, the half-life changed to 16.38 and 15.69 s, where as the experimental half-life is 11.304 s.
Similarly, for $^{30}$P, both USDB and SDNN interactions demonstrated a slight change. Initially, the half-life without EC was 3.040 and 3.261 min for USDB and SDNN interactions, respectively. Upon including EC, the half-life become 3.036 and 3.057 min in both cases, while observed half-life of $^{30}$P is 2.498 min. In $^{34}$Cl, without EC, the half-life is 30.21 min for USDB and 27.63 min for SDNN interaction. With the inclusion of the EC phase space factor, the half-life became 29.89 and 27.35 min, respectively, whereas, the experimental half-life of $^{34}$Cl is 31.99 min. 
Similarly, for $^{38}$K, the half-life without EC inclusion is 3.885 and 3.684 min for USDB and SDNN, respectively. Incorporating the EC phase space factor the half-life would become 3.868 and 3.668 min for USDB and SDNN interaction, respectively, while the observed half-life is 7.651 min, a significant difference is found between theoretical and experimental half-life.
 The EC phase space factor contributions can improve the half-life if we considered the additional corrections arise from the screening of the nuclear charge by atomic electrons and from the finite nuclear size in Eq. \ref{eq:13}.
In general, the EC phase space factor influences the decay half-life of some nuclei more significantly than others due to the variations in energy available for the decay ($Q$-value), atomic orbital density, imperfect atomic wave function overlap in the initial and final state of decay, environmental factors such as temperature, pressure, etc. which tends to change the electron density, type of decay, relativistic effect which comes into play at higher atomic number Z $>$ 70. It is also depends on the forbiddeness of the decay, in the present study we only consider allowed transitions so this effect can be ignore.

%%%%%%%%%%%%%%%%%%%%%%%%%%%%%
 The SM results of $logft$ values, branching ratios, and half-lives of $^{25}$Si, and $^{26}$P in $sd$ space using USD interaction are reported in ref. \cite{Thomas}. In \cite{Thomas}, The theoretical half-lives of $^{25}$Si, and $^{26}$P are reported as 216.1 and 40.85 ms respectively, while the measured values are 220(3) and 43.7(6) ms respectively. In the present work, the half-lives using USDB and SDNN interactions are 238.7 and 218.7 ms for $^{25}$Si and 43.13 and 42.11 s for $^{26}$P. Our results for half-lives of $^{26}$P are more closer to the experimental data than previous SM results, while for $^{25}$Si the SDNN prediction is more closer to the observed value.

 The half-life of $^{22}$Al has been calculated using two different $Q$-values reported in \cite{Achouri} and \cite{audi}. A significant difference have been observed when using different $Q$-values. When using a $Q$-values of 18.601 MeV \cite{audi} a quenched half-life of 45 and 41.67 ms have been obtained for USDB and SDNN interactions, respectively. However when a $Q$-values of 17.56 MeV \cite{Achouri} is used, a significant improvement have been seen in both USDB and SDNN interactions with half-life of 56.34 and 50.71 ms, respectively.

A superallowed transitions with $0^+$ $\rightarrow$ $0^+$ for $^{36}$Ca observed at 4281.9 keV with branching ratio 38.03$\%$. The calculated Fermi matrix element ($M_{F}$) for the transition $0^+$ $\rightarrow$ $0^+$ are 4.04$\times$$10^{-5}$ using USDB and 4$\times$$10^{-5}$ using SDNN interactions respectively, and the corresponding Fermi reduced transition probability ($B(F)$) are 1.63$\times$$10^{-9}$ and 1.6$\times$$10^{-9}$. Due to small Fermi reduced transition probability the half-life contribution from superallowed transition is negligible and hence a significant difference between theory and experiment are observed for $^{36}$Ca.

%%%%%%%%%%%%%%%%%%%%%%%%%%%%%%%%%%%%%%%%%%%%%%%%%%%%%%%%%%%%
\begin{figure*}
\begin{center}
\resizebox{1.0\textwidth}{0.40\textwidth}{\includegraphics{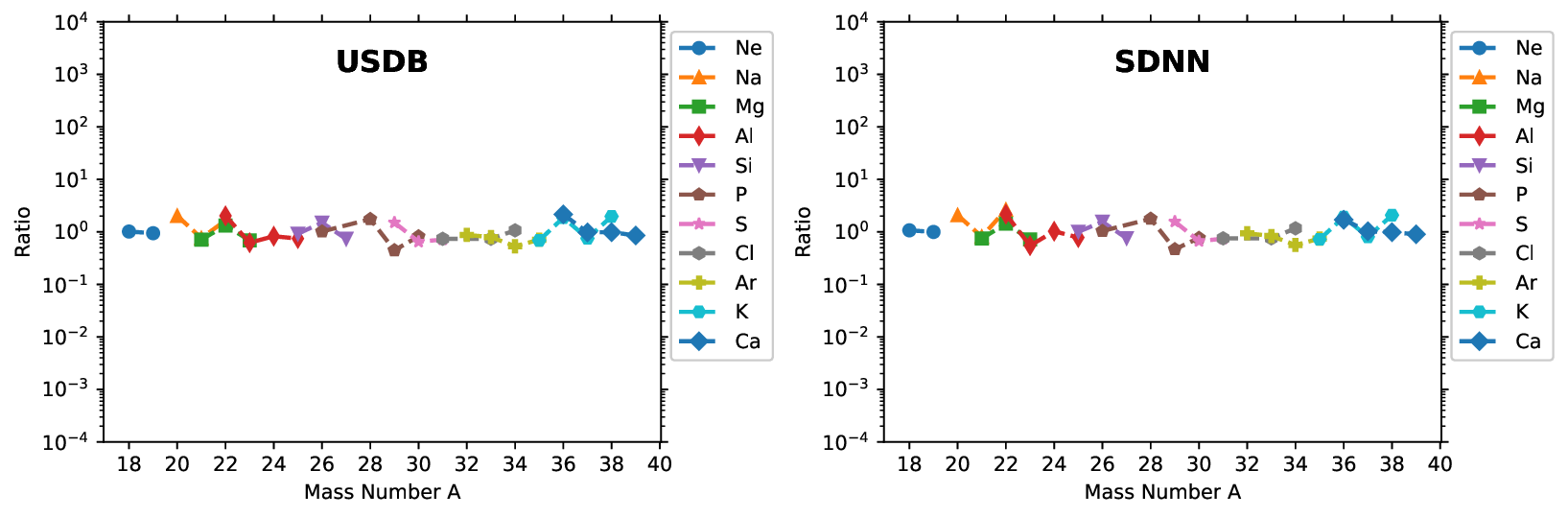}}
\caption{\label{fig3} The theoretical to experimental half-lives ratios versus the mass number $A$ for the concerned $sd$ shell nuclei.}
\end{center}
\end{figure*}
%%%%%%%%%%%%%%%%%%%%%%%%%%%%%%%%%%%%%%%%%%%%%%%%%%%%%%%%%%%%
\begin{figure*}
\begin{center}
\resizebox{1.0\textwidth}{1.2\textwidth}{\includegraphics{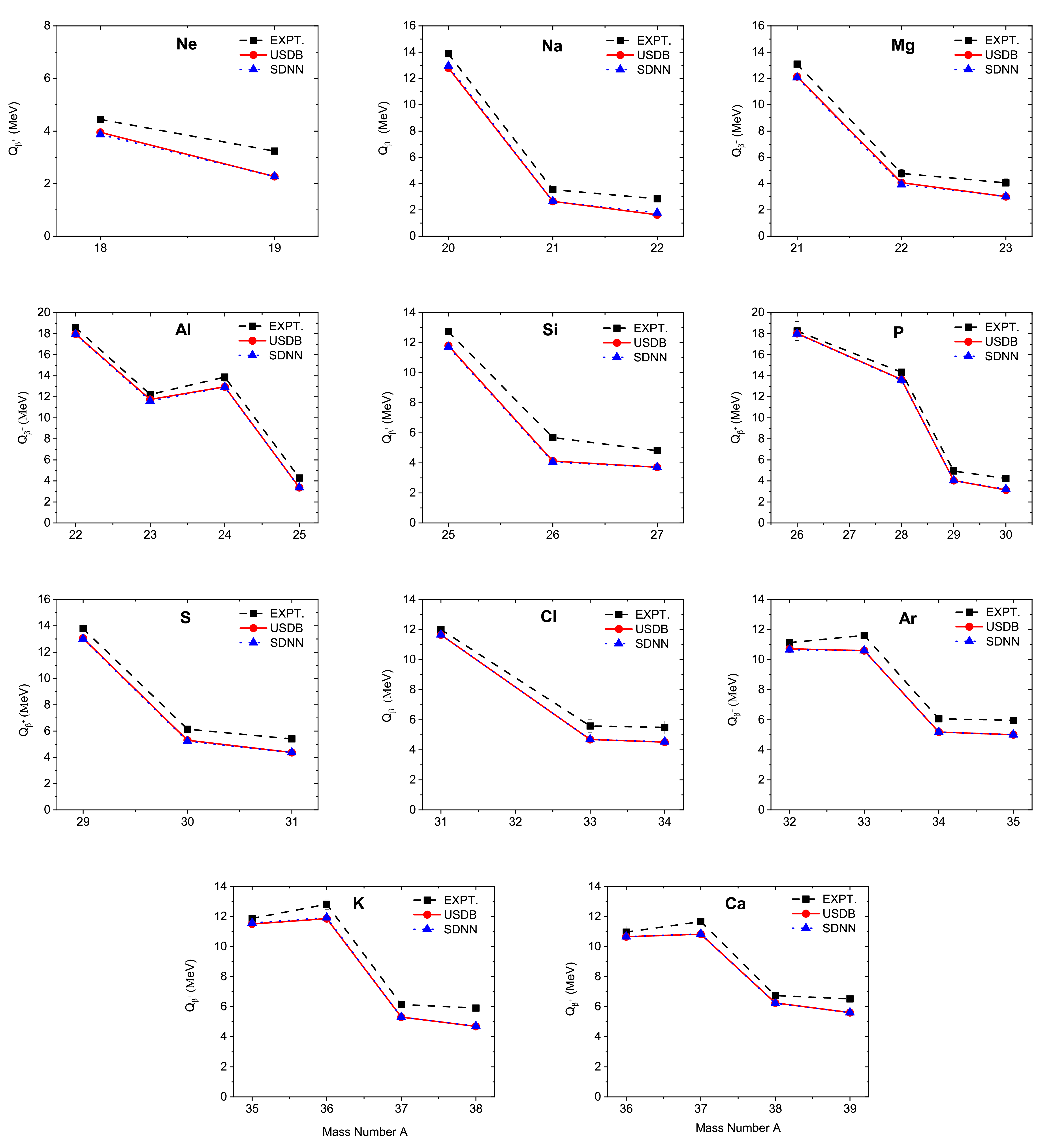}}
\caption{\label{fig4} The theoretical and experimental $Q$ values versus mass number A of the concerned $sd$ shell nuclei.}
\end{center}
\end{figure*}

%In the cases of $^{23}$Mg, $^{29}$P, and $^{30}$P, the contribution of the EC phase space factor is quite less.

%This is because our calculations are unable to predict the ground state correctly, and uncertainties of the
%$Q$ values are large. These facts are responsible for the large errors of the theoretical half-lives.

The ratios between theoretical $\beta^+$-decay half-lives and experimental data for the concerned nuclei are plotted in Fig.\ref{fig3}. For the majority of the isotopes, the ratios between theoretical $\beta^+$-decay half-lives and experimental values are distributed quite close to unity. 

Fig. \ref{fig4} shows $\beta^+$-decay $Q$ -values between theoretical results using Eq.~\ref{eq:16} and the experimental data from ref. \cite{ENSDF}. The figure shows a recurring pattern: as the mass number of a particular isotope increases, the associated $Q$ -value decreases. Moving towards lower $Q$ -values in beta decay signifies an approach to more stable nuclei. The $Q$ -value behaviour for each $\beta^+$-decay exhibits a consistent pattern in both interactions, which resembles the experimental outcomes. Moreover, the $Q$ -value for each $\beta^+$ -decay is slightly lower than the experimental data. The rms deviation of $Q$ -value between theory and the experiment is 880.4 keV, and the average relative error is 13.55$\%$ for USDB interaction, while for SDNN interaction, the rms deviation is 886.4 keV with 13.6$\%$ average relative error. For Al and Ca, the relative error of the theoretical $Q$ -values is less than 10$\%$, and for Ne and Na, it is greater than 20$\%$, and for the rest of the nuclei (Mg, Si, P, S, Cl, Ar and K), the relative error lies in between 10 - 20$\%$.  
%\newpage
\section{Summary and Conclusion}
In the present work, we have reported a systematic SM study of $\beta^+$/EC -decay half-lives, log$ft$ values, branching fractions and $Q$ -values.
%for 37 proton-rich and $N = Z$ nuclei of $sd$ shell. 
The calculations have been performed in full $sd$ model space using two different USDB and SDNN interactions. We have also reported the quenching factor for $sd$ space using USDB and SDNN interactions with Gamow Teller matrix elements for 116 decays.  Further, the $\beta^+$/EC -decay properties are calculated with axial-vector coupling constant $g_A$$(= 1.27)$, and $\kappa$ value $(= 6289)$. We have also included superallowed $0^{+}$ $\rightarrow$ $0^{+}$ transition for seven nuclei and the electron phase space factor for the required nuclei in $\beta$-decay half-lives calculations. Overall, the calculated results of excitation energies, log$ft$ values, quenched half-lives, and branching fractions are in reasonable agreement with the available experimental data. 
The present shell model results of $\beta^+$/EC -decay half-lives, log$ft$ values, branching ratios, $Q$ -values for the $sd$ shell nuclei will add more information to the earlier experimental works, and the calculated average quenching factors in this work can be used for half-life calculations in $sd$ shell.

 %%%%%%%%%%%%%%%%%%%%%%%%%%%%%%%%%%%%%%%%%%%%%%%%%%%%%%%%%%
 \newpage
\section*{Acknowledgment:}
Surender acknowledges financial support from BHU for his PhD thesis work. V. Kumar acknowledges financial support from IoE Seed Grant BHU \\
(R/Dev/D/IoE/Seed Grant-II/2021-22/39960), and SERB Project (File No. EEQ/2023/000157), Govt. of India.

\bibliography{utphys}
\bibliography{references}

\end{document}